\documentclass[letterpaper,useAMS,usenatbib]{mn2e}
\usepackage{comment}
\usepackage{amssymb}   
\usepackage{amsmath}   
\usepackage{amssymb}   
\usepackage{bm}

\usepackage{xspace}
\usepackage{graphicx}
\usepackage{verbatim}

\usepackage[T1]{fontenc}
\usepackage{aecompl}

\newcommand{\be}{\begin{equation}}
\newcommand{\ee}{\end{equation}}

\title[Narrow-line flux ratios in HE0435]{Probing dark matter substructure in the gravitational lens HE0435-1223 with the WFC3 grism}
\author[Nierenberg et al.]{
A.~M.~Nierenberg$^{1, 2}$\thanks{\tt nierenberg.1@osu.edu}, 
T.~Treu$^{3,4}$, 
G.~Brammer$^{5}$,
A.~H.~G.~Peter$^{1,6,7}$,
C.~D.~Fassnacht$^{8}$,
\and
~C.~R.~Keeton$^{9}$, 
~C.~S.~Kochanek$^{1,6}$, 
K.~B.~Schmidt$^{10}$,
D.~Sluse$^{11}$,
S.~A.~Wright$^{12}$
\\
\medskip\\
$^1$ Center for Cosmology and AstroParticle Physics, 191 West Woodruff Avenue, The Ohio State University, Columbus OH 43204, USA \\
$^2$ CCAPP Fellow \\ 
$^3$ UCLA Physics \& Astronomy, 475 Portola Plaza, Los Angeles, CA 90095-1547, USA\\
$^4$ Packard Fellow\\
$^5$ Space Telescope Science Institute, 3700 San Martin Dr, Baltimore MD 21211, USA \\
$^6$  Department of Astronomy, The Ohio State University, 4055 McPherson Laboratory, 140 West 18th Avenue, Columbus OH\\
$^7$ Department of Physics, 191 West Woodruff Avenue, The Ohio State University, Columbus OH 43204, USA \\
$^8$  UC Davis Department of Physics, 1 Shields Ave., Davis CA 95616 \\
$^9$  Department of Astrophysics \& Astronomy, Rutgers, 36 Frelinghuysen Rd, Piscataway, NJ 08854 \\
$^{10}$  Leibniz-Institut f\"{u}r Astrophysik Postdam, An der Sternwarte 16, 14482 Potsdam, Germany \\
$^{11}$  University of Li\`{e}ge, Department of Astrophysics, Geophysics, Oceanography, All\'{e}e du 6 A\^{o}ut, 17, B5C, 4000 Sart Tilman, Belgium \\
$^{12}$ UCSD Center for Astrophysics and Space Science, 9500 Gilman Dr, La Jolla CA 92093-0424\\
}

\begin{document}
\date{Accepted for publication in MNRAS}
\pagerange{\pageref{firstpage}--\pageref{lastpage}}\pubyear{2016}

\maketitle           

\label{firstpage}

                      
\begin{abstract}
Strong gravitational lensing provides a powerful test of Cold Dark Matter (CDM) as it enables the detection and mass measurement of low mass haloes even if they do not contain baryons. Compact lensed sources such as Active Galactic Nuclei (AGN) are particularly sensitive to perturbing subhalos, but their use as a test of CDM has been limited by the small number of systems which have significant radio emission which is extended enough avoid significant lensing by stars in the plane of the lens galaxy, and red enough to be minimally affected by differential dust extinction.  Narrow-line emission is a promising alternative as it is also extended and, unlike radio, detectable in virtually all optically selected AGN lenses. We present first results from a WFC3 grism narrow-line survey of lensed quasars, for the quadruply lensed AGN HE0435-1223.  Using a forward modelling pipeline which enables us to robustly account for spatial blending, we measure the [OIII] 5007 \AA~ flux ratios of the four images. We find that the [OIII] fluxes and positions are well fit by a simple smooth mass model for the main lens. Our data rule out a $M_{600}>10^{8} (10^{7.2}) M_\odot$ NFW perturber projected within $\sim$1\farcs0 (0\farcs1) arcseconds of each of the lensed images, where $M_{600}$ is the perturber mass within its central 600 pc. The non-detection is broadly consistent with the expectations of $\Lambda$CDM for a single system. The sensitivity achieved demonstrates that powerful limits on the nature of dark matter can be obtained with the analysis of $\sim20$ narrow-line lenses.

\end{abstract}

\begin{keywords}
dark matter -- 
galaxies: dwarf --
galaxies: haloes --
quasars: individual (HE0435-1223)
\end{keywords}
\setcounter{footnote}{1}

\section{Introduction}
\label{sec:intro}

A key prediction of Cold Dark Matter (CDM) is that the halo mass function should follow an unbroken power-law with $dN/dM \propto M^{-\alpha}$, where $\alpha=1.9\pm0.1$ from cluster down to planet mass halos \citep{Diemand++08, Springel++08}. CDM models of the distribution and growth of structure match observations with remarkable success over an enormous range of distance and size scales \citep[e.g.][]{Planck++13}. On smaller scales, tests of CDM become more difficult owing to the uncertain physics of star formation in sub-Milky Way mass halos ($M_{vir}<10^{12} M_\odot$). A famous example of this is the `Missing Satellite Problem', so called due to the fact that CDM simulations predict that thousands of subhalos should be gravitationally bound to the Milky Way, while only $\sim$ tens of \emph{luminous} satellite galaxies have been observed  \citep{Moore++1999, Klypin++99, Strigari++07, Weinberg++08, Drlica-Wagner++15}. Extrapolations based on the depth and area of the Sloan Digital Sky Survey (SDSS) and the Dark Energy Survey (DES) forecast that the Milky Way may host as many as $\sim$100 ultra-faint ($L<10^3 L_\odot$) satellite galaxies \citep[see, e.g.][]{Hargis++14, Drlica-Wagner++15}. However, measurements of the halo mass function at these low luminosities becomes extremely difficult. In particular, the stars of low luminosity galaxies occupy only the inner $\sim100$ pc of their dark matter halo. This is a small fraction of the total halo virial radius, which, in absence of the effects of tidal stripping extends out to several kpc even for a very low mass $10^{7}M_\odot$ halo.  One consequence of this is that satellite galaxies inhabit halos with kinematically consistent masses over 5 orders of magnitude in luminosity \citep{Strigari++08}. 

Simulations with varying implementations of semi-analytic and/or numerical baryonic feedback have demonstrated that it is possible to suppress star formation in dark matter subhalos sufficiently to matched the observed Milky Way satellite luminosity function with an underlying CDM halo mass function at low redshift \citep[e.g.,][]{Thoul++1996,Gnedin++00,Kaufmann++08, Maccio++11, Springel++10,Guo++11, Zolotov++12, Brooks++13, Starkenburg++13,Weinberg++08, Menci++14, Lu++14, Wetzel++16}. However, models which can match the luminosity function of satellites at redshift zero around the Milky Way, are not always successful at reproducing the luminosity function around higher mass hosts, or at higher redshifts \citep[e.g.][]{Nierenberg++13, Nierenberg++16}. 

Strong gravitational lensing provides a powerful test of CDM as it enables a measurement of the subhalo mass function without requiring stars or gas to detect subhalos. In a strong gravitational lens, a background source is multiply imaged with the image positions and magnifications depending on the first and second derivatives of the gravitational potential, respectively. The image magnifications are particularly sensitive to low-mass perturbations, with a lower sensitivity limit determined by the source size. If the source is of the scale of $\mu$as (such as a quasar accretion disk), then the image magnifications will be significantly affected by stars in the plane of the lens galaxy (a.k.a microlensing). In contrast, milliarcsecond scale sources are not significantly lensed by stars, but are sensitive to the presence of perturbing subhalos which have characteristic Einstein radii of $\sim$mas, corresponding to typical masses $10^4<M/M_\odot<10^9 $ given typical lens configurations.

Traditionally, radio loud quasar sources have been used to detect subhalos, as they are extended enough (tens to hundreds of parsecs, e.g. \citet{Jackson++15}) to avoid microlensing, and also are not affected by differential dust extinction in the plane of the lens galaxy. \citet{Dalal++02} used the lensed magnification of 6 radio loud quasar lenses and PG1115+08 \citep{Weyman++80} to estimate the average fraction of mass in substructure relative to the mass in the smooth halo component ($f_{sub}$) around these lenses, finding this fraction to be broadly consistent with predictions from CDM although with large uncertainties. This was an important proof of method. Progress requires a larger sample of gravitational lenses which enables a measurement of not only the fraction of mass in substructure but also the slope of the subhalo mass function. Furthermore, \citet{Xu++15}, \citet{Hsueh++16}, and \citet{Gilman++16} have demonstrated that systematic uncertainties may occur in flux-ratio measurements if the deflector is not accurately modelled due to insufficiently deep optical imaging, as may be the case for studies which rely entirely on radio imaging.


There are several paths forward for increasing the sample of lenses which can be used to measure the subhalo mass function. Gravitational imaging for instance, can be used to detect subhalos perturbing the lensed positions of background galaxies. This method currently has a limiting mass sensitivity of $\sim 10^8 M_\odot$ \citep{Vegetti++12,Vegetti++14, Hezaveh++16, Birrer++17}.  The sensitivity limit is determined in part by imaging spatial resolution, which enables the measurement of the small astrometric perturbations to the strongly lensed backgrounds source caused by substructure. The next generation of telescopes and adaptive optics will lower this limit for background galaxy sources, and deep VLBI imaging of extended radio jets will potentially enable the detection of masses as low as $\sim 10^6 M_\odot$ with gravitational imaging. 

Observations of quasar lenses at longer wavelengths can provide a microlensing-free probe of substructure. Redward of $\sim 4 \mu$m rest-frame, the quasar continuum emission is expected to be dominated by the dusty torus, which is sufficiently large to be unlensed by stars \citep[e.g.,][]{Sluse++12}. For a typical source redshift, this implies that mid-IR imaging at wavelengths greater than $10\mu$m can probe dark matter substructure. This method has been applied successfully to several systems with larger image separations \citep[e.g.,][]{Macleod++09,Minezaki++09, Chiba++05}. JWST can provide the spatial resolution required to extend mid-IR flux ratio measurements to systems with smaller image separations.  \citet{Jackson++15} demonstrated that deep radio observations of radio-quiet lensed quasars can also successfully be used as a microlensing free probe of substructure, albeit with larger flux uncertainties than radio-loud systems ($\sim 8-10\%$ compared with $\sim 3-5\%$ for radio loud systems). They estimate that this method can be applied to approximately half of optically selected quasar lenses.

Strongly-lensed quasar \emph{narrow-line} emission provides an alternate probe of substructure, with comparable precision to radio-loud lensing studies. This method, originally proposed by \citet{Moustakas++03}, is extremely promising as it enables the measurement of substructure with current observational facilities in virtually all of the tens of optically selected quadruple quasar lenses predicted to be found in DES \citep[][{\tt http://strides.astro.ucla.edu/}]{Ostrovski++16, Agnello++15} and other wide field imaging surveys, and the hundreds forecast to be found in LSST \citep{Oguri++10}. \citet{Sugai++07} demonstrated this method for the gravitational lens RXS 1131-1231 using seeing-limited observations with the Subaru integral field spectrograph. This lens has an unusually large separation of $>1$ arcsecond between each of the images. 

Higher resolution imaging is necessary for the majority of quad lens systems, which often have at least one pair of images separated only by a few tenths of an arcsecond. Adaptive optics can provide the necessary spatial resolution to isolate the lensed images. \citet[][hereafter N14]{Nierenberg++14} used the integral field spectrograph, OSIRIS \citep{Larkin++06}, at Keck with adaptive optics to measure spatially resolved narrow-line flux ratios in the gravitational lens B1422+231 \citep{Patnaik++92}. Adaptive optics is an effective tool, however it can only be applied to those systems in which the narrow-line emission of interest falls in a suitable wavelength range for adaptive optics correction, for instance either $H$ or $K$ band in the case of Keck OSIRIS. Furthermore, adaptive-optics requires the presence of a nearby, bright, tip-tilt star, although often the lensed quasar itself is bright enough for this purpose. Space-based spatially resolved spectroscopy provides an alternative for those systems which fall outside of these wavelength bands, or are at a declination unsuitable to Keck OSIRIS spectroscopy.
 
In this work we demonstrate that the Hubble Space Telescope infrared grism on the Wide Field Camera 3 (WFC3) can be used to measure spatially-resolved narrow-line image fluxes, with comparable sensitivity to substructure to ground-based results from Keck. We present an analysis of WFC3 grism observations of HE0435-1223 \citep[hereafter HE0435,][]{Wisotzki++02} to demonstrate our reduction mechanism. The deflector is an early-type galaxy at redshift 0.4546 \citep{Morgan++05} and the source is a quasar at redshift 1.693 \citep{Sluse++12}.  This system has been extensively studied since its discovery; it has been monitored over a decade, and its relatively long time delay and significant intrinsic variability have made it a powerful probe of the group density profile and the Hubble constant  \citep{Wong++16, Bonvin++16, Courbin++11, Kochanek++06}. Thanks to this attention, there are numerous multi-band and spectroscopic measurements available for comparison and significant effort has gone into measuring the properties of the environment of the deflector \citep[e.g.]{Momcheva++06, Momcheva++15, Wong++11,Sluse++16, Wilson++16}, which is important for comparing detected subhalo properties with predictions from CDM. 
 
In Section 2 we describe the observations and initial data reduction for this system. In Section 3 we describe the spectral extraction pipeline developed to measure narrow-line fluxes. In Section 4 we report measured quasar spectral features, and integrated emission fluxes, and compare the measured fluxes to results from broad band and radio studies.  In Section 5 we perform a simple gravitational lens inference to test for the presence of substructure. In Section 6, we test for the effects of resolved narrow-line emission on our results. In Section 7 we discuss the constraint from this system. In Section 8 we provide a brief summary of the main conclusions.


We assume a flat $\Lambda$CDM cosmology with
$h=0.7$ and $\Omega_{\rm m}=0.3$.  All magnitudes are given in the AB
system \citep{Oke++1974}.

\section{Observations and Initial Reduction}
\label{sec:data}
We observed the gravitationally lensed quasar HE0435 as a part of HST-GO-13732 (P. I. Nierenberg), a grism survey of narrow-line emission in six quasar lenses.  The target was observed on August 30, 2015 for 2062s with the G141 Grism, and for 400s with F140W direct imaging. The observation was taken at a dispersion angle of 147 degrees East of North so that the dispersed quasar images would be maximally separated from each other along the direction perpendicular to the dispersion axis. In order to recover sub-pixel information, we split the observations into a four point dither pattern with half-integer sub-pixel offsets following the procedure of \citet{Brammer++12} \citep[see also][]{Schmidt++14,Momcheva++16}.
For each dither position, we took a 100s direct exposure with F140W, immediately followed by a 515s G141 exposure. The F140W direct images at each dither position were used to obtain accurate wavelength solutions for each G141 exposure \citep{Brammer++12}.

Raw F140W and G141 exposures were individually processed with {\tt AstroDrizzle} \citep{Gonzaga++12} in order to reject cosmic rays, remove geometric distortion and perform flat-field subtraction \citep{Koekemoer++11, Brammer++12}. The F140W exposures were drizzled onto a 0\farcs06 pixel scale (approximately half the native pixel size). The upper left panel of Figure 1 shows the drizzled F140W image of the gravitational lens, and nearby spiral galaxy G1. 

The G141 exposures were interlaced and combined onto a 0\farcs06 pixel scale, corresponding to an observed wavelength resolution of $\sim 22$~\AA~ per pixel following \citet{Brammer++12}, \citet{Schmidt++14} and \citet{Momcheva++16}. Unlike drizzling, interlacing does not introduce correlated pixel flux errors. 
Figure 1 shows the final interlaced grism data with arrows indicating the 5007 and 4959~\AA~[OIII] doublet which is partially blended given the grism resolution.

\section{Spectral Extraction}
Our goal is to measure the narrow-line emission flux in each lensed quasar image, taking into account blending between distinct spectral components after they are dispersed by the grism. The lower left panel of Figure 1 highlights the blending by showing how the light from the ring, quasar images and galaxies combine in a model grism image. In order to rigorously account for the overlapping spectra in the grism image, we employ a forward modelling approach. We discuss this method in detail in the following subsections; in brief, we generate a model direct image and use the {\tt 3D-HST} grism simulation code \citep{Brammer++12} to iteratively map proposed component spectra into a simulated 2D grism image, which is then compared with the original interlaced grism image to compute a $\chi^2$ goodness of fit. In Section 3.1 we discuss how the model direct image is generated, in 3.2 we discuss the 1D models we use for the spectral components, and in 3.3 we describe the statistical inference.



\subsection{Direct image model}

The grism image is effectively a convolution of an object spectrum with its direct image. Thus given a model for the direct image, it is possible to generate a predicted grism spectrum. We model the direct image as having seven distinct spatial plus spectral components: Four separate quasars, the main deflector, the quasar host galaxy which is lensed into a ring, and the nearby galaxy G1. Here we discuss how the direct image model is generated for each of the components. These direct model images are then combined with model 1D spectra, as discussed in Section 3.2, in order to generate model grism images.

We model the four quasar images as point sources, using a nearby star to model the point spread function (PSF). We optimize the point source positions and fluxes using {\tt galfit} \citep{Peng++02, Peng++10}. A possible concern for using a drizzled star image as the model for the point source is that the true PSF is not accurately captured at the exact location of the lensed images. Furthermore, the FWHM of the grism PSF varies slightly with wavelength, which the {\tt 3D-HST} pipeline does not account for in the forward modelling process. We have checked that the exact PSF model does not impact our inference on the [OIII] flux ratios by running the entire modelling process described in the next two subsections with a total of five different PSF models: 1) A median combination of stars in the F140W FOV, 2) and 3) The median star blurred by 10 and 15\%, 4) and 5) two different nearby stars. While the choice of PSF model affected the overall fit to the 2D grism image, we found that it had no impact on our inference of the relative [OIII] fluxes. The quasar image positions are listed in Table A1, with uncertainties given by the variation in best fit {\tt galfit} positions for the different PSF models.


In order to disentangle the lens galaxy light from the prominent ring and bright quasar images, we start with the empirical model of the deflector by \citet{Wong++16} derived from very deep (9337 s), F160W imaging. This model is generated from a superposition of Chameleon profiles \citep{Dutton++11}, and is based on a simultaneous fitting of the quasar point images, a model for the lensed background quasar host galaxy, and the lens. To generate a model for the F140W light profile, we start by fitting two S\'ersic profiles to the empirical F160W model. Next, we hold all of the parameters for the S\'ersic profiles obtained in the previous step fixed except for the total flux, and fit the F140W direct image as a combination of the lens and quasar images. We subtract the best fit galaxy and QSO models from the direct F140W image and are left with a residual image composed primarily of the ring. In the third step, we subtract the residual ring image from the original F140W image, and re-fit the ring-subtracted image with the galaxy and QSO models, this time allowing all of the S\'ersic parameters for the galaxy model to vary freely. The final model for the galaxy is taken from the third step. We have tested several different iterations of this process, including allowing for less flexibility in the galaxy model to verify that the inferred narrow-line fluxes and image positions are not sensitive to the exact galaxy model, although the overall fit to the 2D grism image varies.

The ring model is generated by subtracting the best fit lens galaxy and quasar models from the direct image. We mask a small region near the centre of each QSO image where the PSF subtraction is noisy. We have confirmed that the inferred [OIII] fluxes are not sensitive to the exact size of the masked region. Finally for G1, we simply use a small cutout of the direct image, which is possible because it is isolated in the direct image.

\begin{figure*}
\centering
\includegraphics[scale=0.6, trim = 0 0 0 0 clip=true]{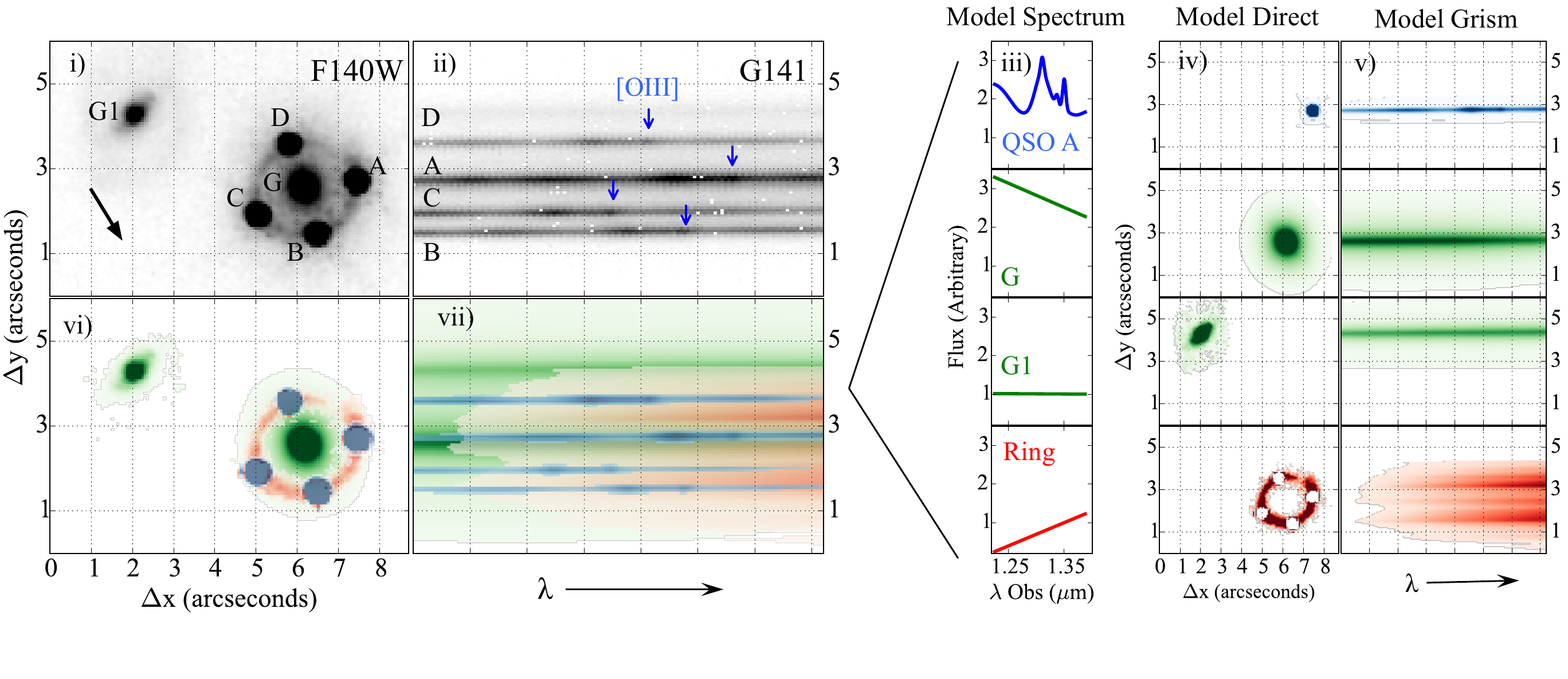} 
\caption{Demonstration of the forward modelling method used to infer spectral parameters. Note that the image contrasts have been altered between images to highlight different features.
{\bf Panel i)} Drizzled F140W image, arrow indicates North. {\bf Panel ii)} Interlaced G141W grism image, with light dispersed along the x-axis of the F140W image. QSO spectra (A-D) are labeled. They overlap with spectra from the ring, the main deflector (G) and the spiral galaxy (G1). Blue arrows indicate the location of narrow [OIII] 4959 and 5007 \AA~emission which are partially blended at this resolution. {\bf Column iii)} MCMC proposed 1D spectra for four of the seven components labelled in panel i. Each of the QSO images A-D has a separate model spectrum (shown in Figure 3), only spectrum A is shown here.
{\bf Column iv)} Model direct images for each separate spectral component, described in Section 3.1. The central QSO pixels are masked in the ring model to account for noisy PSF subtraction in this region.
{\bf Column v)} Model 2D grism images for each spectral component generated from convolving the model spectra in column iii with the model direct image in column iv. 
{\bf Panels vi, vii)} Final, combined model direct image and model grism image, generated from the sum of columns iv and v respectively (and the other three QSO images not shown). Colours are the same as in columns iii, iv and v. 
The goodness of fit is calculated by the $\chi^2$ difference between true and model 2D G141 images.}

\label{fig:prmass}
\end{figure*}

\subsection{1D Spectral Models}

 In this section we describe the analytic models we use for the 1D spectra. Although our grism data extend over a significantly larger wavelength region, we confine our model and comparison with the data to a small wavelength region around [OIII]. Figure 2 shows the fitted regions for each QSO image. The extent of the region is approximately rest frame 4500-5200~\AA, but it varies between the lensed QSO images and is selected to achieve two goals. First, to provide sufficient spectral coverage to obtain a good constraint on the broad Fe, H$\beta$ and continuum features which overlap with the [OIII] emission (Figure 2). Second, the modeled region is extended where there is a possibility of an emission feature blending with the [OIII] emission of a neighbouring 2D spectrum. An example of the latter case is found in the redward extension of the fitted region for QSO spectrum C, in order to include the possible broad Fe flux contribution from image C to the image B narrow line emission (Figure 1).  

Although gravitational lensing is not wavelength-dependent, the spectra of the four QSO images may not necessarily be related by a simple multiplicative magnification factor owing to the intrinsic variability of the QSO (which can lead to different image spectra owing to the varying image arrival times), and to the differential effect of stellar microlensing as a function of intrinsic source size. Thus we construct a model for the four QSO spectra which enables variations due to these effects. The QSO spectrum in the wavelength range of interest is composed of broad Fe and H$\beta$ emission, continuum emission and narrow [OIII] and H$\beta$ emission. 

We model the broad H$\beta$ emission as a Gaussian with an intrinsic redshift offset from the [OIII] emission left as a free parameter. Although other studies have found that the H$\beta$ line profile can have a complicated structure, the low resolution of the grism data does not warrant a higher order model; Figure 2 demonstrates that a Gaussian is sufficient to fit the line profile in this case. The broad Fe emission is modelled using IZw1 templates which we interpolate in velocity space following \citet{Bennert++11} and \citet{Woo++06}. The broad Fe emission velocity is independent of the broad emission width in our model. The model allows for a redshift offset between the broad Fe emission and the [OIII] emission. 

We model the continuum emission as a straight line rather than a power-law, given the small fractional size of the wavelength region of interest. The slope and amplitude of the line are left as free parameters.
The broad-line and continuum emission both arise from regions which can be affected by microlensing, which we discuss in more detail in Section 4. To account for this possibility we allow the width and amplitude of the H$\beta$ emission and the slope and amplitude of the continuum emission to vary independently between the model spectra. We did not find significant evidence for variations in the broad Fe velocities between images, and so kept them fixed for our final analysis, however we allowed the Fe amplitudes to vary independently from the H$\beta$ amplitudes. 

Unlike the broad and continuum emission, narrow [OIII] and H$\beta$ emission come from a sufficiently extended source (greater than tens of parsecs) to not be affected by either stellar microlensing or intrinsic variability \citep{Moustakas++03, Muller-Sanchez++11, Bennert++06a, Bennert++06b}. Owing to this, we assume that both the line widths and the relative amplitudes of [OIII] and narrow H$\beta$ should be constant between the lensed images. We model the [OIII] doublet and H$\beta$ narrow-lines as Gaussians, and assume that they have the same redshift, which is valid given the spectral resolution of the grism. The ratio of the [OIII] doublet 4959 and 5007 amplitudes is fixed to the quantum-mechanically predicted value of 1/3.

The 1D models for the deflector, ring and G1 spectra are modelled as straight lines over the short wavelength region of interest, with amplitudes and slopes as free parameters. We do not find evidence requiring the inclusion of emission or absorption features in any of these spectra relative to the measurement uncertainties and given the brightness of the QSO spectra (see e.g. Figure 2). 

We assume that the image fluxes are not affected by differential dust extinction. In the rest frame of the lens, the [OIII] emission lines lie at  roughly $\sim9300$~\AA. At this wavelength, total dust extinction in lens galaxies, and early-type galaxies in general, is typically of order only a few hudredths of a magnitude \citep[e.g][]{Fal++99, Ferrari++99}, which is well within our overall flux measurement uncertainty. This assumption is further supported by the similarity of the broad-band optical colours of the images \citep{Wisotzki++03}. The images also have mutually consistent CIV (lens rest frame $\sim 2790$ \AA) and H$\beta$ (lens rest frame $\sim 9300$~\AA) broad-line flux ratios.


\subsection{Inference of QSO spectral parameters}

We infer the probability distribution of the parameters of the 1D spectral models using a Bayesian forward modelling approach with the {\tt emcee} Markov Chain Monte Carlo software package \citep{Foreman-Mackey++13}. For each step, the MCMC algorithm proposes parameters for the 1D spectra of all seven distinct spectral components (four QSO images, the main galaxy, the lens ring and G1).
We then simulate dispersed images of each separate component and add them to generate a full model 2D grism image. Finally, the $\chi^2$ of the fit is computed relative to the original 2D interlaced image. Figure 1 illustrates how the model 2D direct image components are dispersed into the model 2D grism image for each MCMC step.

\begin{figure*}
\centering
\includegraphics[scale=0.55]{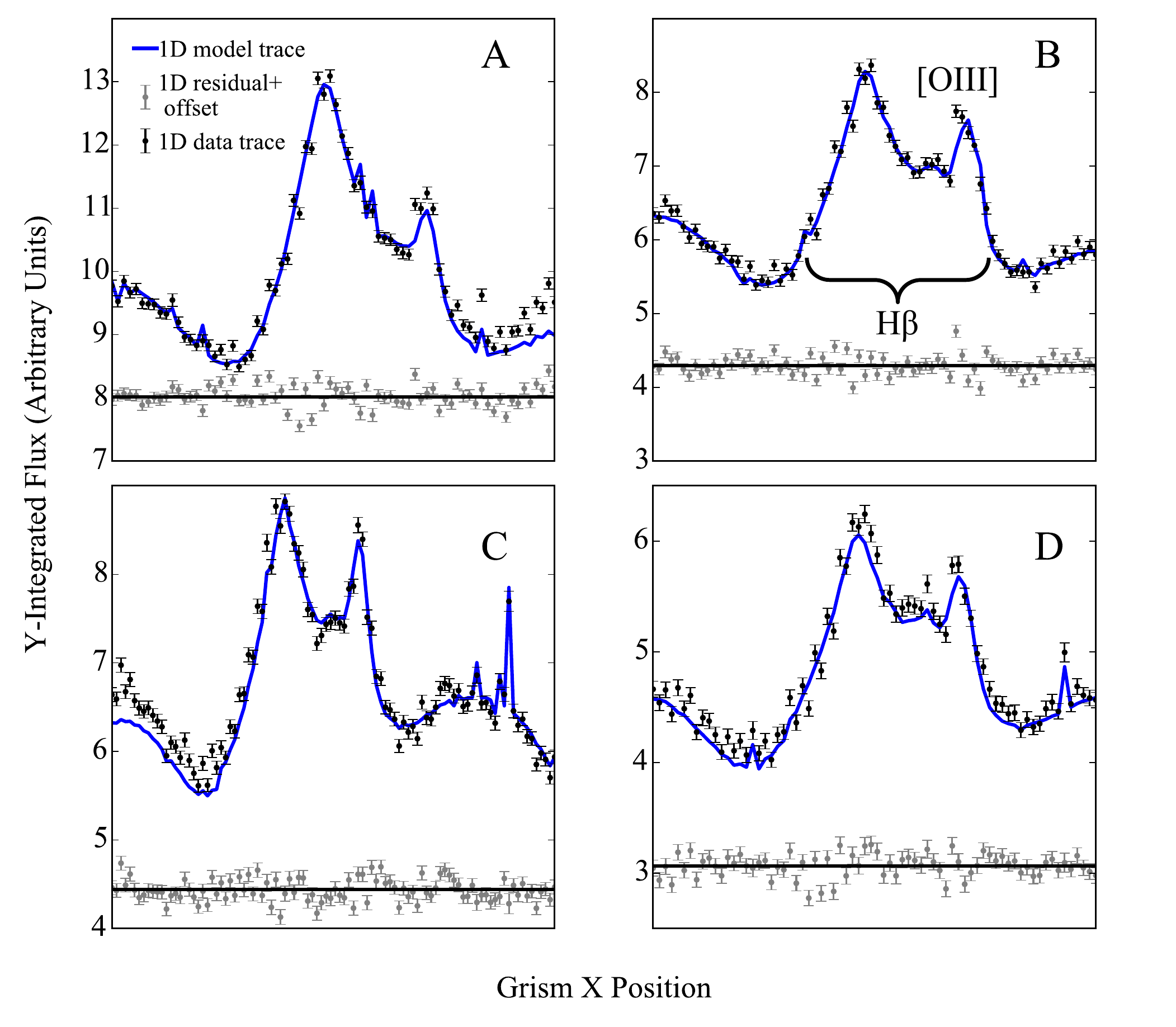}
\caption{Lensed quasar spectra extracted along the x-axis of the 2D grism image (Figure 1) via PSF weighted averaging along the y-axis. Absolute fluxes are arbitrary. 1D Spectra include contamination from neighbouring dispersed QSO images, lens galaxy light, the lens ring, and the nearby spiral galaxy G1 as illustrated in Figure 1. The residual is derived by subtracting the 2D grism model from the 2D grism data, and then performing the PSF weighted y-axis averaging; thus it is not a simple subtraction of the blue line from the black points. The residual has been offset from zero by the amount indicated by the horizontal lines for ease of visualization. The modelled region varies slightly for each image depending on its position in the 2D image, as discussed in Section 3.2. }
\label{fig:prmass}
\end{figure*}

\section{Spectral Forward Modelling Results}
Figure 2 shows the 1D model, data and residual `traces' for the four lensed QSO images. These traces are obtained by integrating the flux along the y axis in the 2D image, weighted by the relative flux of the direct F140 model PSF along that axis. Jumps in flux are due to small misalignments between the dispersion axis and the detector axis. This comparison shows that the input model provides an excellent fit to the observed spectra. 

From the spectral modelling we obtain flux ratios between the broad H$\beta$ fluxes and the [OIII] fluxes from the image pairs A/C, B/C, D/C.  Given that the intrinsic quasar luminosity is not known, gravitational lensing analyses rely on ratios of image fluxes rather than their absolute values. In Figure 3, we compare these flux ratios with measurements from other studies across a range of filters and for fixed filters at multiple dates. These measurements are chosen to represent how the flux ratios vary with wavelength and time, and are only a small subset of the many measurements of this system obtained for time variability studies \citep[e.g.][]{Bonvin++16, Courbin++11, Kochanek++06}. Table A2 contains references and observing dates for all flux ratios plotted in Figure 3. 

The narrow [OIII] flux ratios are strikingly different from optical to near-IR flux ratios which are subject to contamination by microlensing and intrinsic QSO time variability.

HE0435 has been monitored for 15 years \citep{Bonvin++16, Courbin++11, Kochanek++06}, and during that time has shown highly variable broad band flux ratios due to stellar microlensing and intrinsic variability. The intrinsic variability particularly affects images B and D which have time delays of over a week relative to images A and C. Figure 3 highlights several repeat measurements of the system which show significant variability. 

Based on simulations of QSO accretion disks and dusty tori, blueward of rest-frame $\sim$4$\mu$m, (observed $\sim$10 $\mu$m) the accretion disk makes a dominant contribution to the QSO emission \citep{Sluse++13}. From chromatic microlensing studies of this system, the quasar  continuum emission has a half light radius of $\sim10^{16.3\pm0.3}$ cm (or 0.003 pc) at a rest-frame wavelength of 8000~\AA. \citet{Sluse++12} estimate the MgII broad line region size to be $\sim10^{18\pm 1}$ cm (or 0.03 pc). These sizes correspond to $0.5-5\mu$as at the lens redshift, and are thus affected by stellar microlensing. \citet{Bonvin++16} have inferred the approximate amplitude of the observed $R$ band (rest frame $\sim 2500 $\AA) stellar microlensing as a function of time for each of the images since 2003. Figure 4 shows an estimate from \citet{Bonvin++16} for the microlensing effect on the $R$ band flux ratios as a function of time assuming a `true' flux ratio value indicated by the straight lines. 
The amplitude of microlensing depends on the source size 
thus bluer filters are more strongly affected by microlensing 
while redder continuum measurements are less affected. \citet{Blackburne++14} have performed a detailed study of differential microlensing as a function of wavelength for this system and their data are included in Figure 3, and in Table A2.

The broad line emission flux ratios of both H$\beta$ from this study and CIII] and CIV from \citet{Wisotzki++03} are closer to the narrow-line emission flux ratios, which is consistent with microlensing being a function of emission region size.

We can test for the effects of microlensing on our data by comparing the relative amplitudes of emission features in our inferred spectra. In Figure 5, we plot the marginalised models for the lensed image spectra from our analysis, normalised to the peak of the [OIII] flux at 5007 \AA~ in order to highlight how the emission features vary 
between the lensed images. Image A shows significant morphological differences, with continuum and broad H$\beta$ fluxes which are much higher relative to the [OIII] flux than the other three images. This indicates that there is significant source-size dependent lensing. This finding is consistent with the inferred $R$ band microlensing of image A as observed by \citet{Bonvin++16} and shown in Figure 4.

The narrow-line [OIII] flux ratios are consistent with 5 GHz radio measurements from \citet{Jackson++15}, with A/C, B/C and D/C differing at 0.25, 1.8 and 2.2 $\sigma$ respectively. This is expected given that both sources are expected to be extended enough to avoid all microlensing contamination.
Although the results do not differ significantly, we note that \citet{Jackson++15} found that their radio emission was somewhat resolved, with an intrinsic source size of $\sigma\sim 288 $ pc, assuming the source had a Gaussian flux distribution. This affects the flux ratios predicted from gravitational lensing relative to a point source for a fixed deflector mass model. We discuss this further in Section 6, where we also place limits on the size of the narrow-emission region in our data and we examine the effects of a resolved narrow emission line region on our results.

\begin{figure*}
\centering
\includegraphics[scale = 0.7]{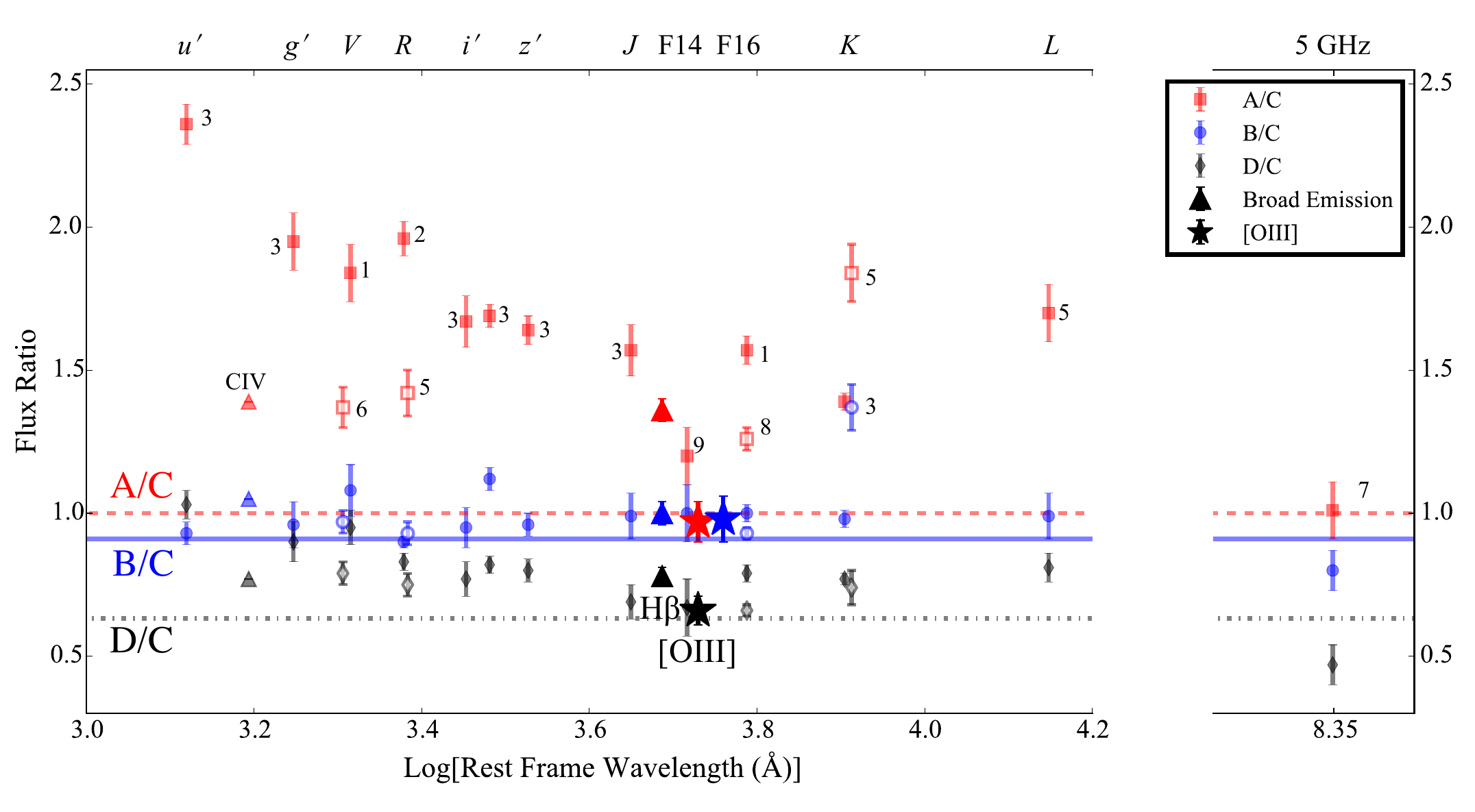}
\caption{Flux ratio measurements for HE0435-1223 selected to represent variations with wavelength and time.  References along with measurement dates are listed in Table A2. Numbers correspond to dates plotted in Figure 4. Squares, circles and diamonds indicate broadband continuum flux measurements, which are subject to time-delay induced variability as well as microlensing blueward of $\sim4\mu$m rest frame. Stars and triangles represent  [OIII] and broad-line flux ratios.  Measurements in the same filter, but from different years are slightly offset from each other for clarity, with the later measurement plotted with an open symbol. [OIII] values have been shifted redward to avoid overlap with broad H$\beta$ results. The B/C [OIII] flux ratio has been artificially shifted redward so it does not lie on top of the A/C value. Dashed, solid and dash-dot lines represent the best smooth model prediction for the A/C, B/C and D/C flux ratios respectively, given the image positions and narrow-line fluxes measured in this work. Top labels list observed bands, where F14 and F16 are abbreviations for the HST filters F140W and F160W respectively. }
\label{fig:prmass}
\end{figure*}

\begin{figure}
\centering
\includegraphics[scale = 0.45]{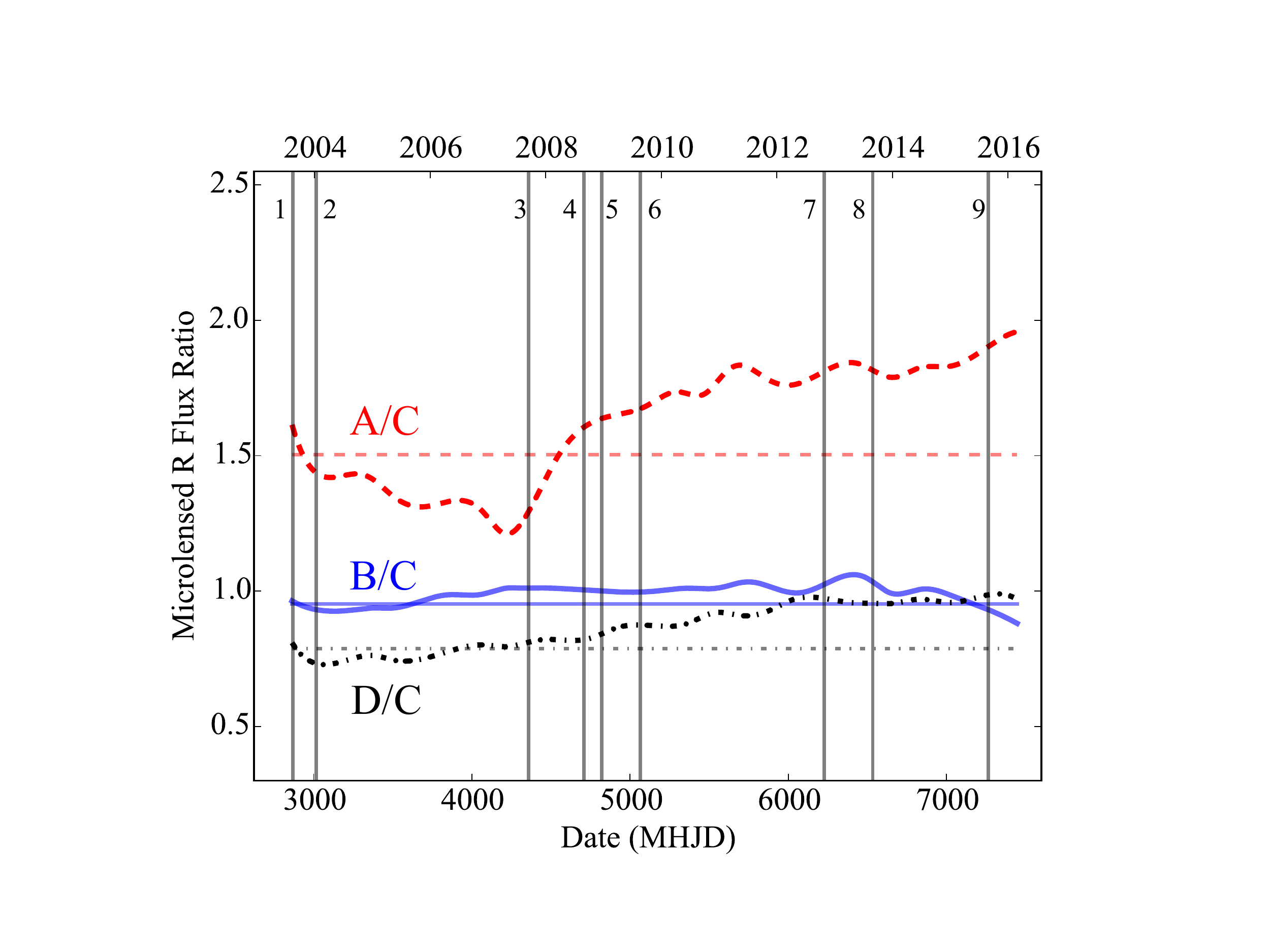}
\caption{An estimate of fluctuations induced by microlensing in the $R$ band flux ratios as a function of time for HE0435, by \citet{Bonvin++16}. Vertical bars with numbers correspond to approximate dates for the measurements plotted in Figure 3 and listed in Table A2. Measurements within 2 months of each other are combined to the same time point. Horizontal lines indicate the model `true' flux ratios at the start of the monitoring campaign. MHJD = HJD-2450000.}
\label{fig:prmass}
\end{figure}

\begin{figure}
\centering
\includegraphics[scale=0.45]{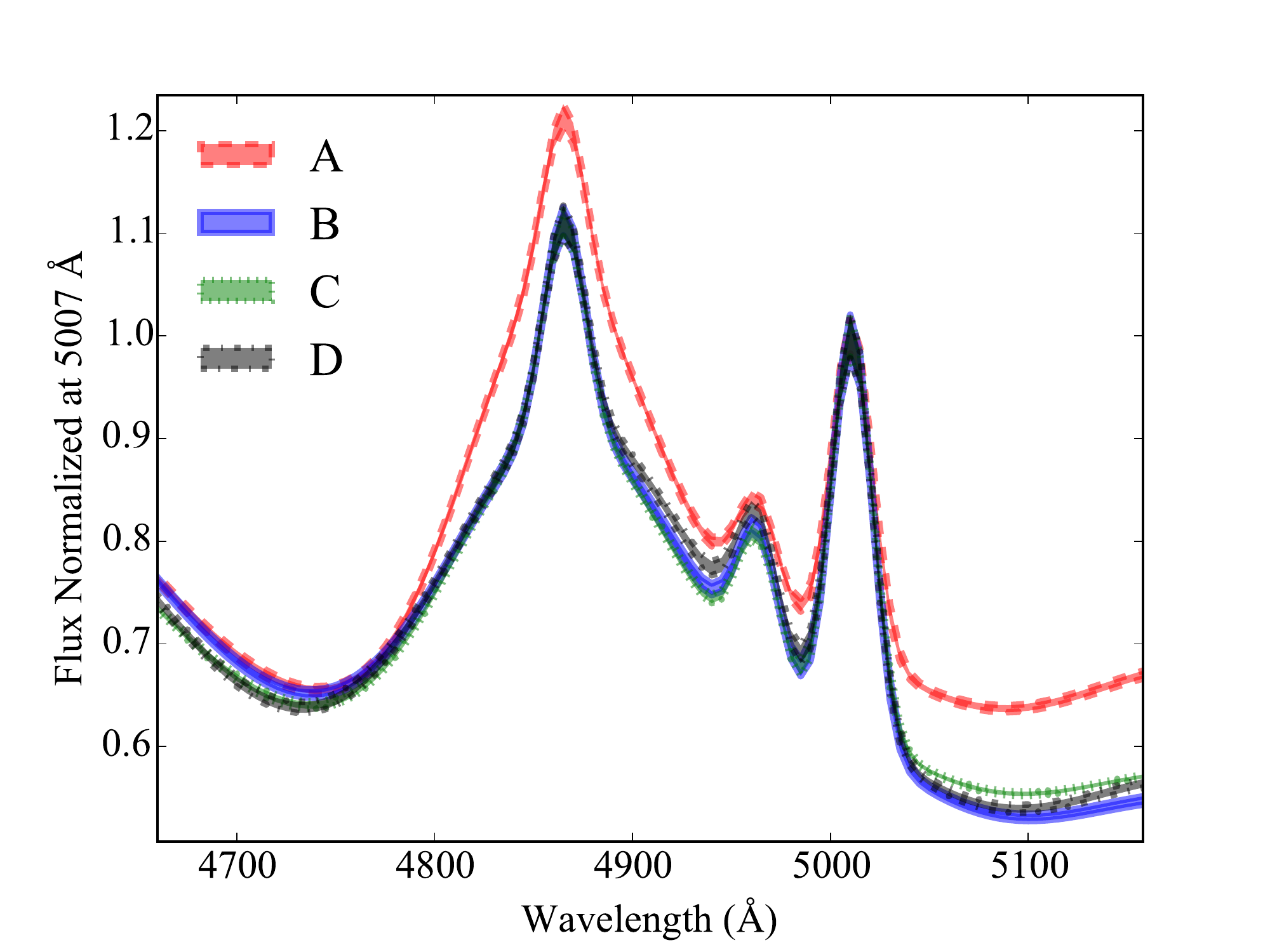}
\caption{Posterior model spectra for each lensed quasar image, normalized to the peak of the narrow-line flux at 5007 \AA~ to highlight variations in emission regions relative to the narrow-line. Image A shows the most significant difference, with continuum and broad line fluxes higher relative to the [OIII] flux than in the other three images.}
\label{fig:prmass}
\end{figure}

\section{Gravitational lens modelling}

The lensed image positions and [OIII] flux ratios are sensitive to the mass distribution of the deflector. As discussed in the Introduction, the image fluxes are particularly sensitive to small scale perturbations caused by dark matter subhalos. In this section we perform a gravitational lensing analysis of the system using the image positions and [OIII] fluxes.
We do not include the time delays as a constraint, given that they are minimally sensitive to perturbers near the lensed images, relative to the image fluxes and positions, and given the time delay measurement uncertainties for this system \citep{Keeton++09a, Keeton++09b}. Furthermore, unlike image positions and fluxes, which are most sensitive to the local mass distribution and can thus be well matched without including G1 explicitly \citep[e.g.][]{Sluse++12}, the time delays are sensitive to the larger scale environment and using them as a constraint would require including the complex multi-plane lensing effect of G1 which is at a higher redshift than the main deflector \citep{Wong++16, Bonvin++16, Jackson++15, Sluse++16}. The macromodel parameters including the external shear term are left free to vary, and thus absorb the large-scale contributions from G1.

In Subsection 5.1 we discuss the optimum smooth mass model fit to the data. In Subsection 5.2 we place limits on the presence of substructure near the lensed images. In Subsection 5.3 we discuss the effects of a finite narrow-line emission region.

\subsection{Smooth Model}
We start with a simple, smooth mass distribution model of the system. We model the main deflector as a singular isothermal ($\rho(r)\propto r^{-2}$) ellipsoid (SIE) which has been shown to provide an excellent match to the combined stellar and dark matter mass distributions of elliptical galaxies well beyond the Einstein radius \citep[e.g.][]{Rusin++03, Rusin++05, Gavazzi++07, Treu++10, Gilman++16}. The SIE has five free parameters; the centroid, the ellipticity and position angle, and the Einstein radius.  The smooth model also includes parameters describing the magnitude and direction of external shear which can be generated by the group environment of the galaxy. 

We optimize these model parameters relative to the observed lensed image positions and [OIII] fluxes using {\tt gravlens} \citep{Keeton++01a, Keeton++01b}, and find an overall best fit $\chi^2$ of 1.7 for one degree of freedom. Table 1 lists the mean and one sigma uncertainties for the lens model parameters. Figure 3 shows the best-fit model fluxes as straight lines. The $\chi^2$ of 1.7 for one degree of freedom indicates that the data has a roughly $\sim 20\%$ chance of being drawn from the best fit model, thus we do not find significant evidence indicating that a more complex model with a significant subhalo contribution is necessary to fit the data.  Our inferred lens model parameters are comparable to other values in the literature which are inferred from microlensing-free data. In particular, our inferred Einstein radius and external shear parameters are consistent with the single lens model parameters from \citet{Sluse++12}. As expected, our inferred spherical equivalent Einstein radius of 1\farcs200$\pm$0\farcs003  for the single lens model is somewhat higher than other values in the literature which explicitly include G1 and find values for the Einstein radius of the main deflector ranging from 1\farcs07$\pm$0\farcs02 \citep{Jackson++15} to 1\farcs182$\pm$ 0\farcs002 arcseconds \citep{Wong++16}. We also infer a somewhat higher value for the external shear than models which include G1 in addition to an external shear; $\gamma_{ext}=$0.063$\pm$0.007 compared with 0.039$\pm0.004$ and 0.030$\pm0.003$  for \citet{Jackson++15} and \citet{Wong++16} respectively. 

This result differs from the result of \citet{Fadely++12} who found that the $L$ band image fluxes and positions could not be fit with a smooth model although they did not report their best fit lens model parameters.  They note that observed $L $ band corresponds to the rest frame 14000 \AA at the redshift of the quasar, which should have a significant flux contribution from the dusty torus in addition to the accretion disk continuum emission \citep{Sluse++13, Wittkowski++04, Honig++08}. In order to test for possible microlensing of the continuum component of the $L$ emission, they analysed two years worth of monitoring data from \citet{Kochanek++06}. Unluckily over this time scale there was not significant evidence for microlensing induced variability in image A. Longer baseline data over 15 years \citep{Bonvin++16} reveals significant microlensing of image A, as shown in Figure 4, which likely affected $L$ band flux ratios. 

In the following subsection we place limits on the presence of perturbing subhalos near the lensed images.

\begin{table}
\centering
\scriptsize\begin{tabular}{lll}
\hline
Parameter & Value  &Description \\
\hline
$\theta_E$ &   1\farcs200$\pm 0.003$ & Spherical Einstein radius \\
q                  &    $0.91\pm 0.03$ &    b/a           \\
PA               &  $-8\pm 5$ &  Degrees E of N            \\
$\gamma_{ext}$       & $0.063\pm 0.007$         & External shear  amplitude      \\
$\theta_\gamma$  & $-18 \pm 2$   & Direction of external shear (Deg. E of N)  \\

\hline \hline
\end{tabular}
\caption{Gravitational lens model parameters for the main deflector and external shear inferred from image positions and [OIII] fluxes which are given in Tables A1 and A2 respectively.}
\end{table}

\subsection{Limits on the presence of substructure}
Each of the image positions and fluxes provides a local constraint on the presence of small scale structure. In this subsection we test the limits on the presence of a single perturbing subhalo given our [OIII] flux and position measurements. We test the measurement sensitivity to two different perturber masses of $M_{600} \sim10^8 M_\odot $ and $10^7 M_\odot$ where $M_{600}$ is the integrated mass within 600 pc of the centre of the perturber. These masses are chosen to be above and below the limit where the `Missing Satellite Problem' is observed in the Milky Way \citep[e.g.][]{Strigari++08}.

As demonstrated in \citet{Nierenberg++14}, the perturber mass distribution can significantly affect the predicted lensing signal with fixed $M_{600}$. This is due to the fact that shallower mass profiles must have a higher overall normalisation than steeper mass profiles in order to achieve the same interior $M_{600}$ integrated mass. This in turn causes the shallower mass profile to have a longer range impact on the observed image fluxes and magnifications. Here we demonstrate the lensing effect for two mass profiles, a singular isothermal sphere (SIS) which has an $m(r)\propto r^{-2}$, and a Navarro, Frenk and White \citep[NFW][]{NFW++1996} halo, which has a shallow interior profile of $m(r)\propto r^{-1}$, which transitions to a steeper value $m(r)\propto r^{-3}$ outside of a scale radius. We obtain the scale radius from the mass-concentration relation predicted by \citet{Maccio++08} assuming a WMAP5 cosmology \citep{Dunkley++09}. 

For each perturber mass, and for each mass profile, we iteratively place the perturber at a fixed position, re-optimize the smooth model parameters, and compare the new best fit $\chi^2$ with the original $\chi^2$ in the absence of a perturbation. We choose the grid spacing qualitatively to ensure that relevant angular dependences are captured in the $\chi^2$ distribution as a function of position. We find that variations are well captured by a spacing of 0\farcs1 for the $ M_{600} \sim10^8 M_\odot$ mass perturber and 0\farcs01 for the  $\sim10^7 M_\odot$ respectively.

Figure 6 shows the projected two and three sigma exclusion regions ($p<5$\% and $0.3$\%)  for a singular isothermal sphere perturber with Einstein radius of 0\farcs01 and 0\farcs001 respectively. Assuming the perturber is in the plane of the lens galaxy, these Einstein radii correspond to integrated masses of $\sim10^{8.2} $ and $10^{7.2} M_\odot$ respectively within 600 pc of the centre of the perturber, making them comparable to the Milky Way satellites Fornax and Sagittarius \citep{Strigari++08}, albeit with steeper mass profiles. Based on the average minimum radius at which including a perturber results in a model probability which is lower than $0.3\%$, the average exclusion radius is $\sim$0\farcs4 (0\farcs1), 0\farcs3 (0\farcs08), 0\farcs4 (0\farcs09) and 0\farcs3 (0\farcs06) for images A, B, C and D for the 0\farcs01 (0\farcs001) Einstein radius perturber. 

Figure 7 shows the projected two and three sigma exclusion regions for an NFW perturber with scale radius of 1\farcs0 and 0\farcs1 corresponding to integrated masses of $\sim10^8 $ and $10^{7.2} M\odot$ within the central 600 pc of the perturber.  Again, assuming a best fit model probability lower than $0.3\%$ be excluded, the average radius of exclusion is A: 1\farcs2 (0\farcs1), B: 0\farcs3 (0\farcs08), C: 1\farcs1 (0\farcs09), D: 0\farcs8 (0\farcs06) for images A, B, C and D for the 1\farcs0 (0\farcs1) perturber. As expected, the NFW perturber must be further from the lensed images than an SIS perturber with the same mass to avoid significantly perturbing their fluxes and positions relative to the best fitting smooth mass model.

\begin{figure*}
\centering
\includegraphics[scale=0.33,trim = 0 0 0 0 clip=true]{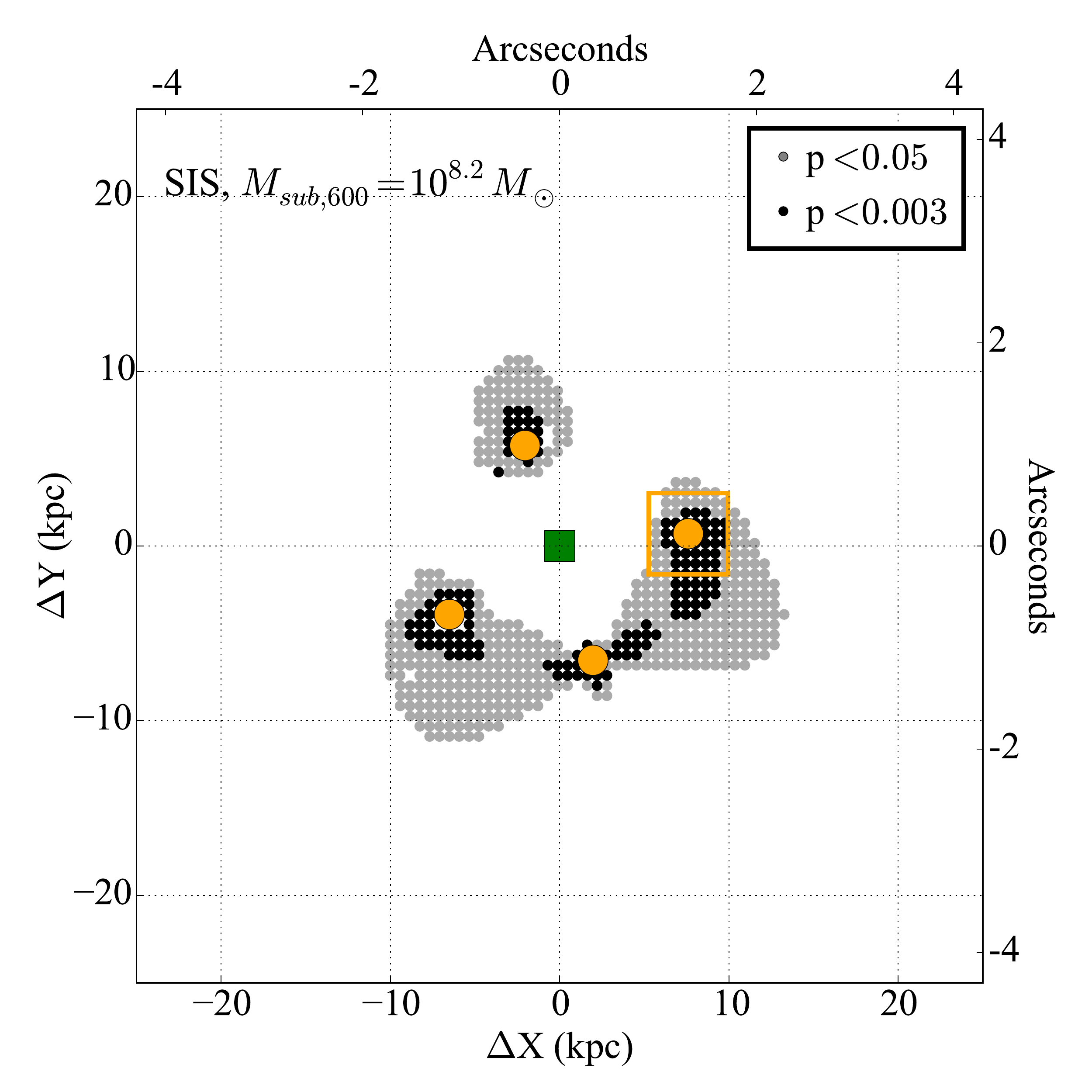} 
\includegraphics[scale=0.33,trim = 0 0 0 0 clip=true]{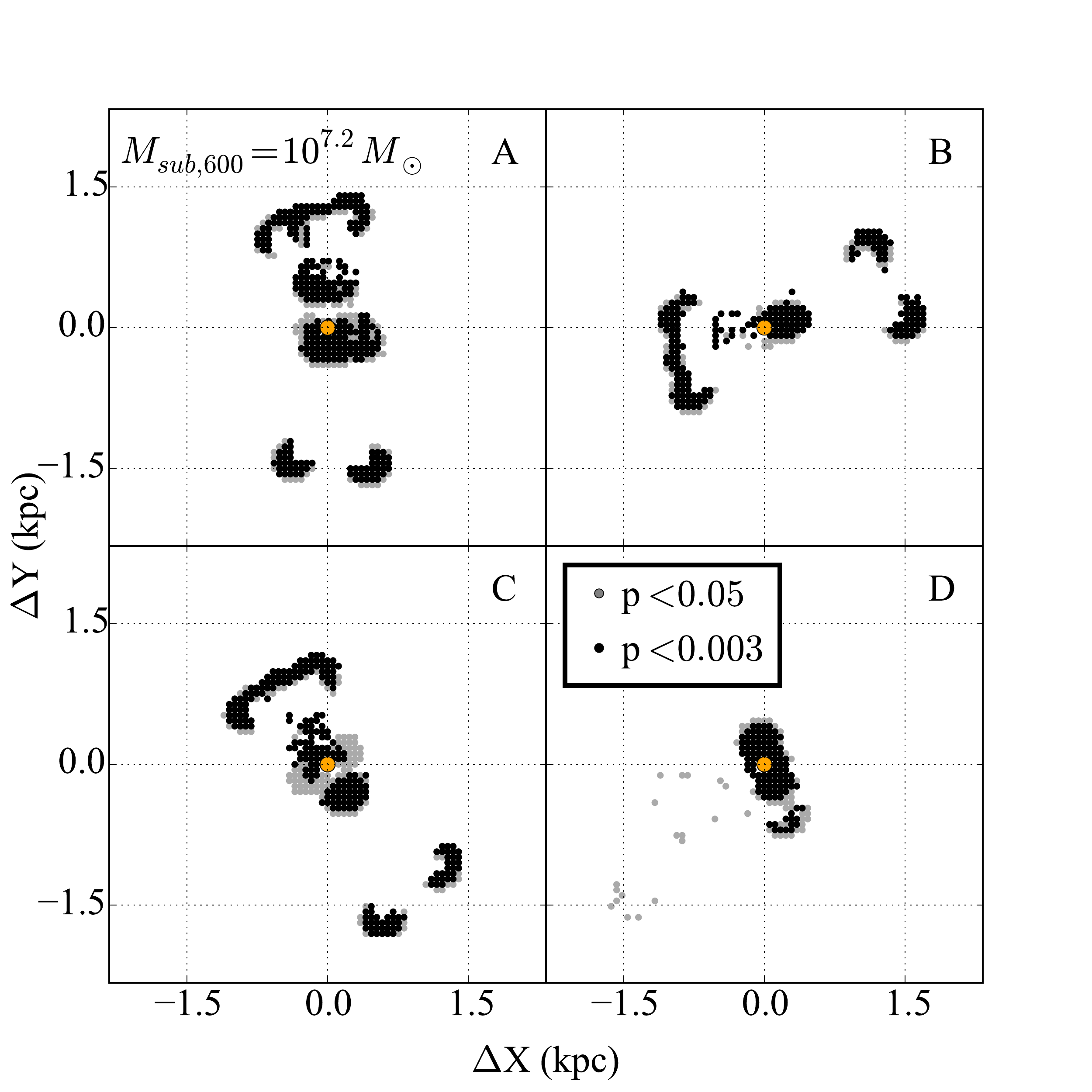} 
\caption{Projected exclusion regions for a singular isothermal spheroid perturber with fixed mass. Light grey and dark grey positions are ruled out with greater than $95\%$ and $99.7\%$ respectively, based on the $\chi^2$ probability of the best fit gravitational lens model to the image positions and [OIII] fluxes after adding a perturber with $M_{600} = 10^{8.2} M_\odot $ (left panel) and $10^{7.2} M_\odot$ (right panel). 
The left panel shows the entire lens system with the green square indicating the lens centroid, and the orange circles representing the quasar images. The orange box in the left panel represents the size of the zoomed regions shown in the right panel. The average projected radial limits are  $\sim$0\farcs4 (0\farcs1), 0\farcs3 (0\farcs08), 0\farcs4 (0\farcs09) and 0\farcs3 (0\farcs06) for images A, B, C and D, respectively, for the $10^{8.2}$ ($10^{7.2} ) M_\odot$ perturber. These exclusion regions correspond to cylinders with radii of $\sim 2 (0.5)$ kpc around each lensed image, projected along the entire host halo. }
\label{fig:prmass}
\end{figure*}

\begin{figure*}
\includegraphics[scale=0.33,trim = 0 0 0 0 clip=true]{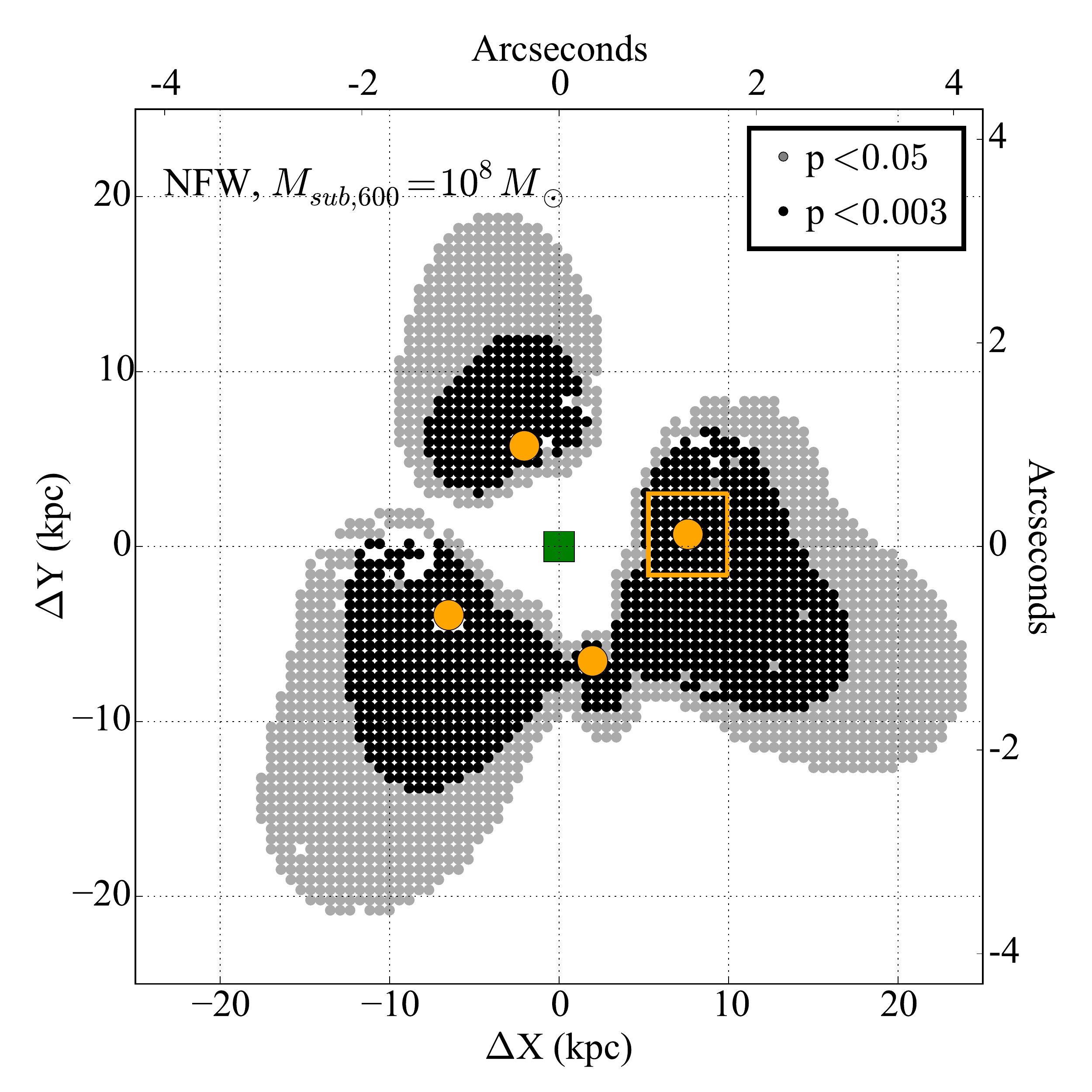} 
\includegraphics[scale=0.33,trim = 0 0 0 0 clip=true]{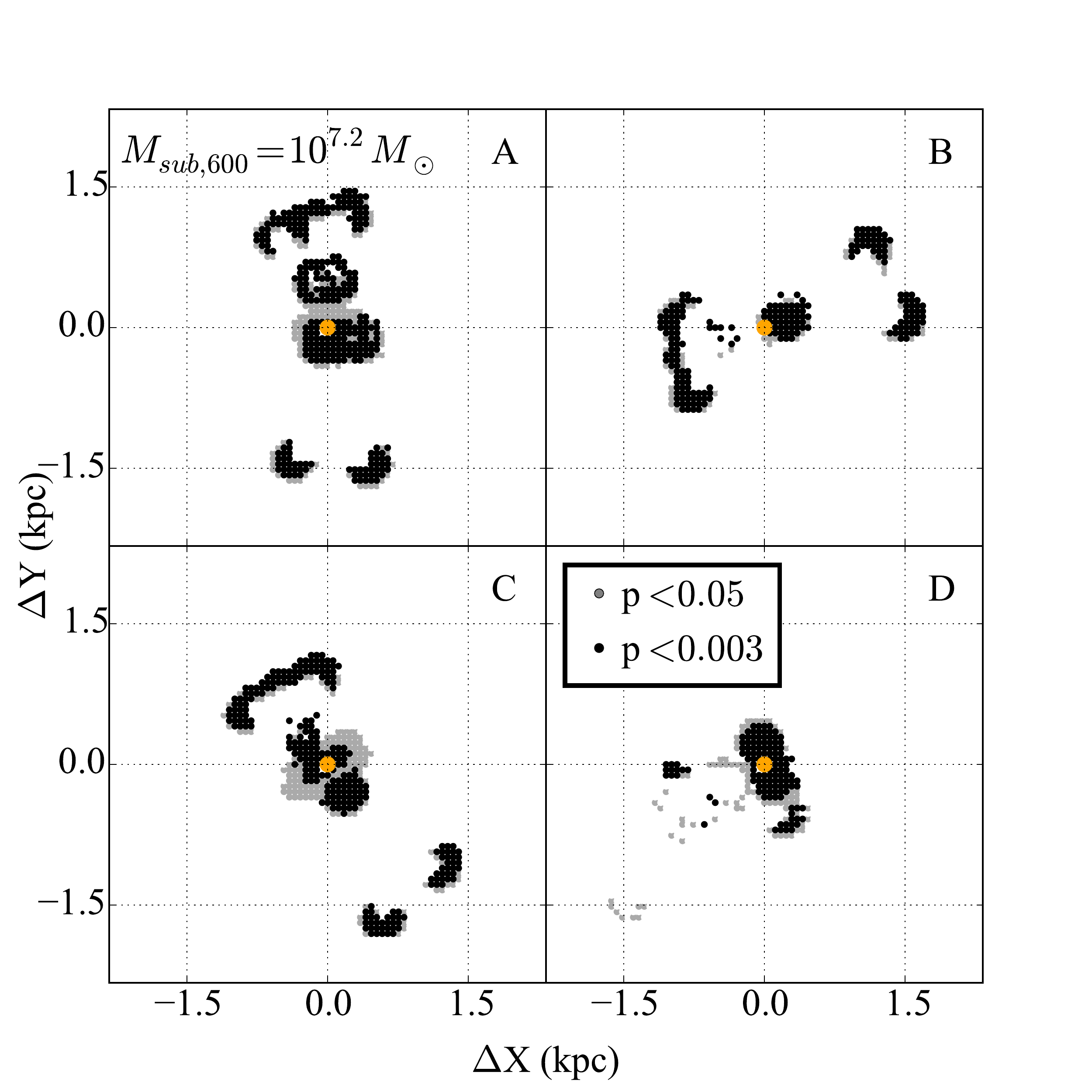} 
\caption{Exclusion regions for an NFW perturbing subhalo with $M_{600} = 10^{8} M_\odot $ and $10^{7.2} M_\odot$, corresponding to NFW scale radii of 1\farcs0 and 0\farcs1 respectively, determined the same way as in Figure 6. The average radial limits are 1\farcs2 (0\farcs1), 0\farcs3 (0\farcs08), 1\farcs1 (0\farcs09), and 0\farcs8 (0\farcs06) for images A, B, C and D, respectively, for a $10^{8}$ ($10^{7.2}) M_\odot$ perturber. These angular scales correspond to an average projected exclusion region of $\sim6$ (0.6) kpc at the redshift of the lens.}
\label{fig:prmass}
\end{figure*}

\section{Finite Source Effects}
Until this point we have assumed that the narrow-line emission in HE0435 is unresolved with HST resolution, even after being strongly lensed and thus magnified. In this section, we explore the effect of a resolved emission region on our results. \citet{Jackson++15} found that the 5 GHz radio emission fluxes for HE0435 were best fit with a smooth deflector mass distribution and a resolved source with 
with an intrinsic size of $\sigma\sim 34$mas (288 pc) for a Gaussian profile. This model significantly improved the fit to their data relative to an unresolved emission model. Given this, it is important that we test whether the narrow-line emission is resolved in our data, and what the effects would be of incorrectly assuming the narrow-line emission was unresolved.

The narrow-line region is extended and in some cases has been observed to extend out several hundreds of parsecs \citep[e.g., ][]{ Bennert++06b}. The narrow-line flux is not uniform, and can be dominated by the central tens of parsecs even for very luminous quasars \citep{Muller-Sanchez++11}. \citet{Sluse++07} demonstrated that the [OIII] emission in the gravitational lens RXS J1131$-$1231 is resolved in their data, and thus inferred a minimum size of $\sim 150$ pc. \citet{Nierenberg++14} found that the narrow [OIII] emission for B1422+231 \citep{Patnaik++92} was marginally resolved with a dispersion of $\sim 10$ mas (50 pc). 

We perform two tests here. First, we simulate narrow-line emission for three different source sizes and then redo the forward modelling inference using these simulated extended narrow-line images in place of the original point source model for the direct image 
to test whether this leads to an improved fit to the observed spectra.
Secondly, we simulate gravitational lenses with three different narrow-line source sizes, and model them under the (in this case, incorrect) assumption that the narrow-emission is unresolved to test for the effect on the inferred flux ratios.

To test the model fit with a resolved narrow-line emission, we repeat the forward modelling inference described in Section 3 with an additional model component. As before, the QSO broad and continuum emission regions are modelled as being emitted by a point source. In order to account for a resolved narrow-line source, we generate a new model direct image for the narrow-line emission only. This extended emission model is generated assuming that the intrinsic source is a Gaussian with width $\sigma_{\rm{NL}}$. We use the best fit gravitational lens model inferred from the continuum image positions and [OIII] flux ratios in order to generate lensed extended emission models for the direct image. We generate models for $\sigma_{\rm{NL}}$ of 50 pc, 100 pc, and 288 pc. The latter source size is the best fit size for the 5 GHz radio emission for this system found by \citet{Jackson++15}. 

We then re-estimate the best fit 1D spectral parameters following the steps in Section 3 with two adjustments. First we use the new simulated extended source model as the direct image model for the narrow-line emission. The other QSO spectral components are modelled as being emitted by point sources as before. Secondly, because the simulated direct image for the [OIII] fluxes by definition fixes the relative image fluxes, the spectral model has only a single parameter for the overall normalization of the narrow-line flux. Thus there are three fewer model parameters than the point source model in which the [OIII] fluxes vary independently. 

The left panels of Figure 8 show the best fit simulated grism images for the narrow-line components only of the three extended models and the point source model for comparison. Note that the narrow H-$\beta$ emission is also modelled as being extended as we assume it is emitted from the same region as the [OIII] emission. While the 50 and 100 pc models differ only marginally from the point source model, the 288 pc source size is clearly extended with HST resolution. The [OIII] doublet for images B and D in particular are nearly completely blended.

 In the right panels of Figure 8 we compare the best fit model 1D trace to the data (analogous to Figure 2) for each of the source sizes. The $\chi^2$ comparison between the data and the simulated grism image grows progressively worse as the source size increases, with best fit $\chi^2$ values of 13228, 13368 and 13990 for 3909 DOF for the 50, 100 and 288 pc narrow-line regions, compared with 13112 for 3906 DOF for the point source model\footnote{The point source model has three extra degrees of freedom as the narrow-line fluxes are allowed to vary independently}. Increasing the narrow-emission source size results in a decreasing best fit peak narrow-line flux, despite the fact that the intrinsic narrow-line emission width is a free model parameter. This is due to the significantly extended emission which cannot be well fit in 2D.

\begin{figure*}
\centering
\includegraphics[scale=0.21,trim = 0 0 0 0 clip=true]{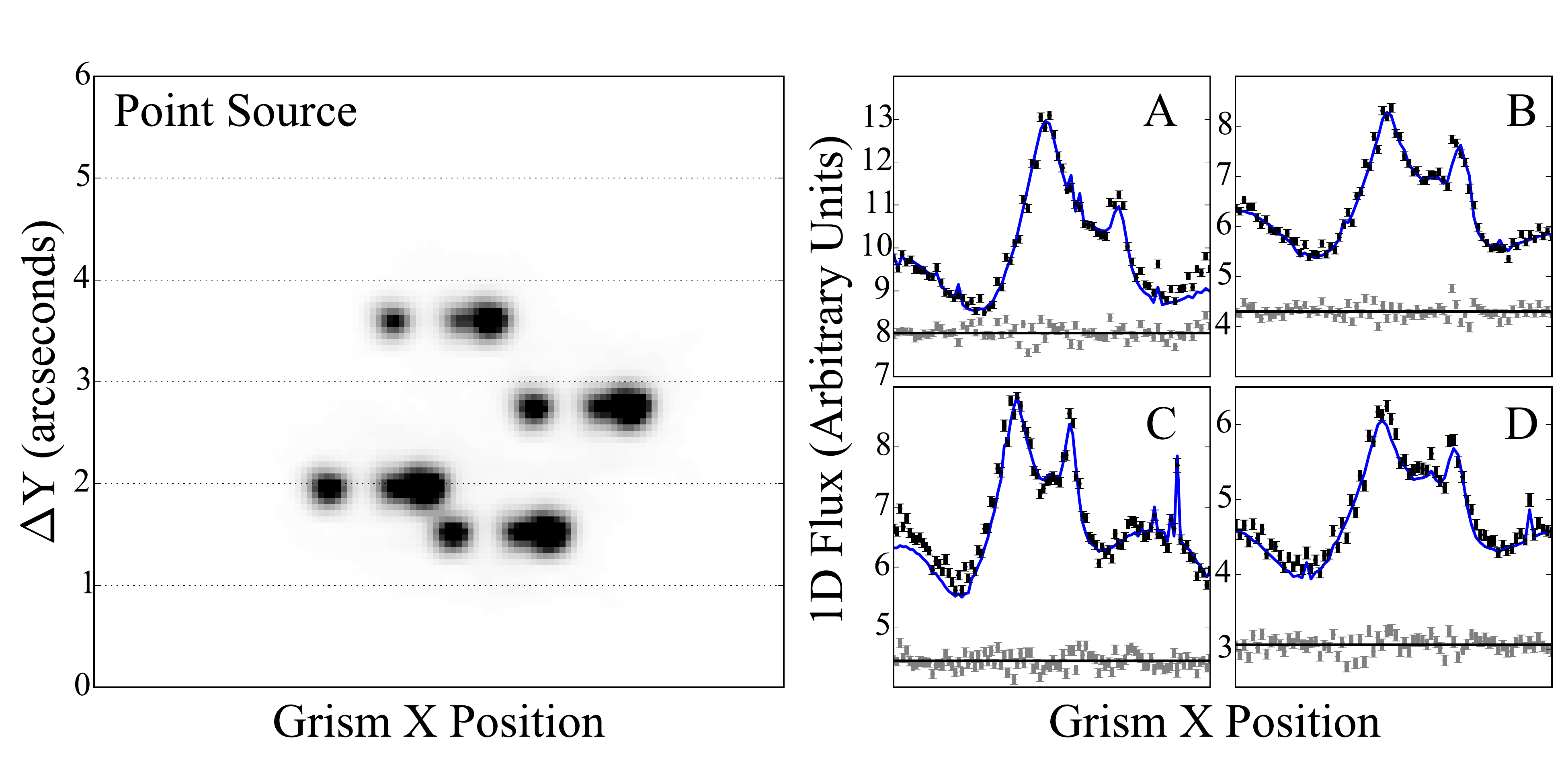} 
\hspace{5mm}
\includegraphics[scale=0.21,trim = 0 0 0 0 clip=true]{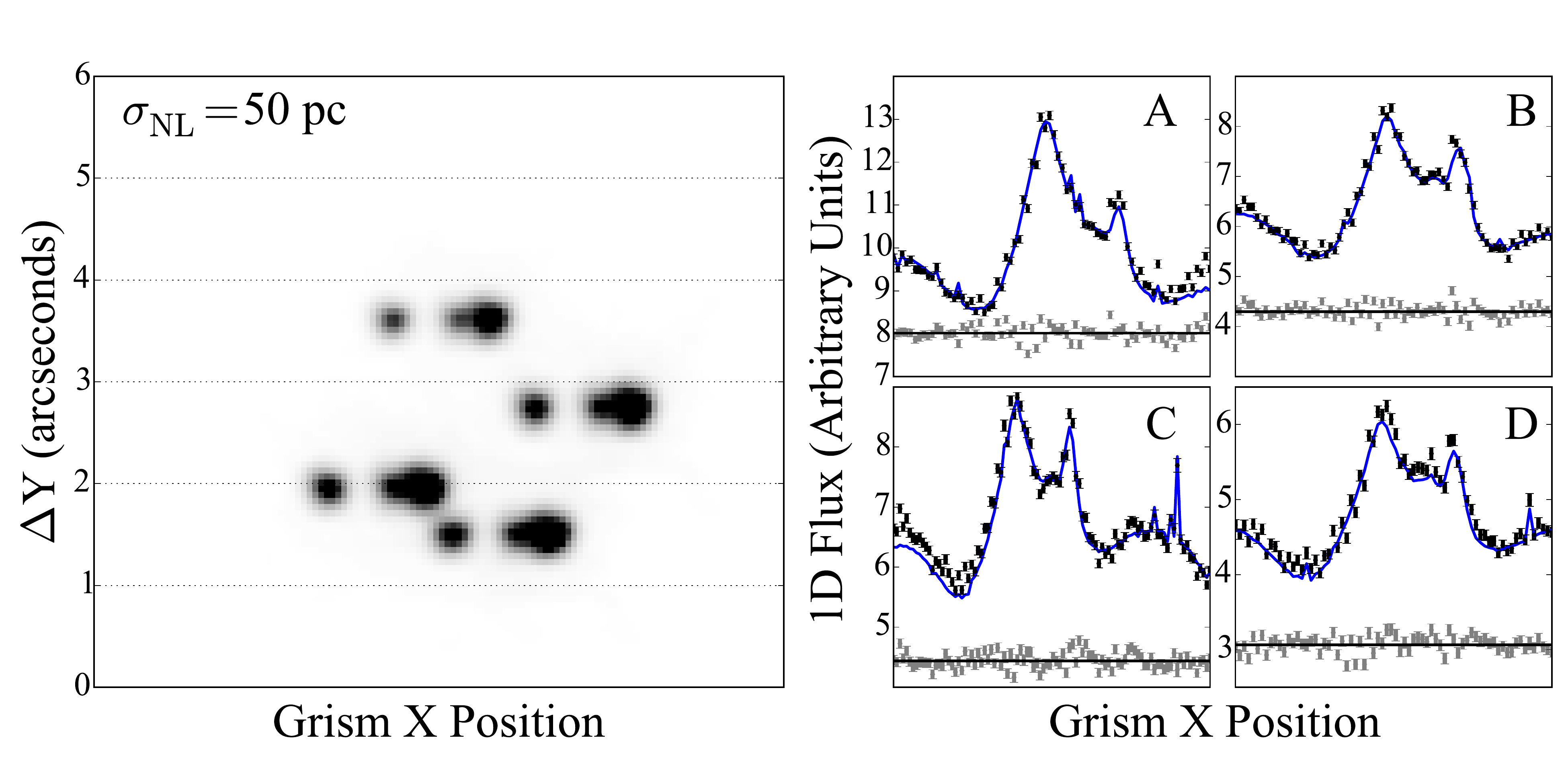} 
\\
\includegraphics[scale=0.21,trim = 0 0 0 0 clip=true]{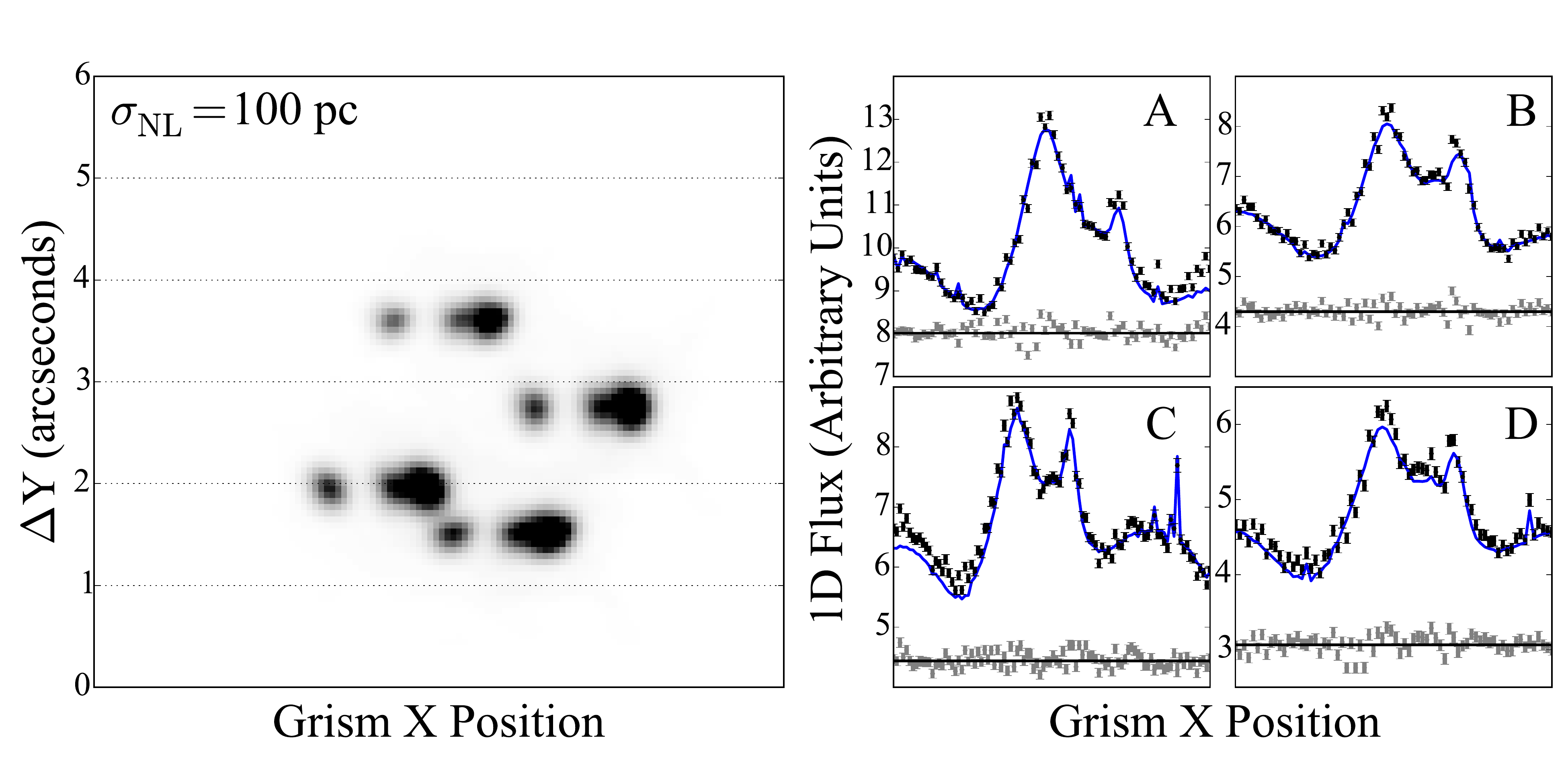} 
\hspace{5mm}
\includegraphics[scale=0.21,trim = 0 0 0 0 clip=true]{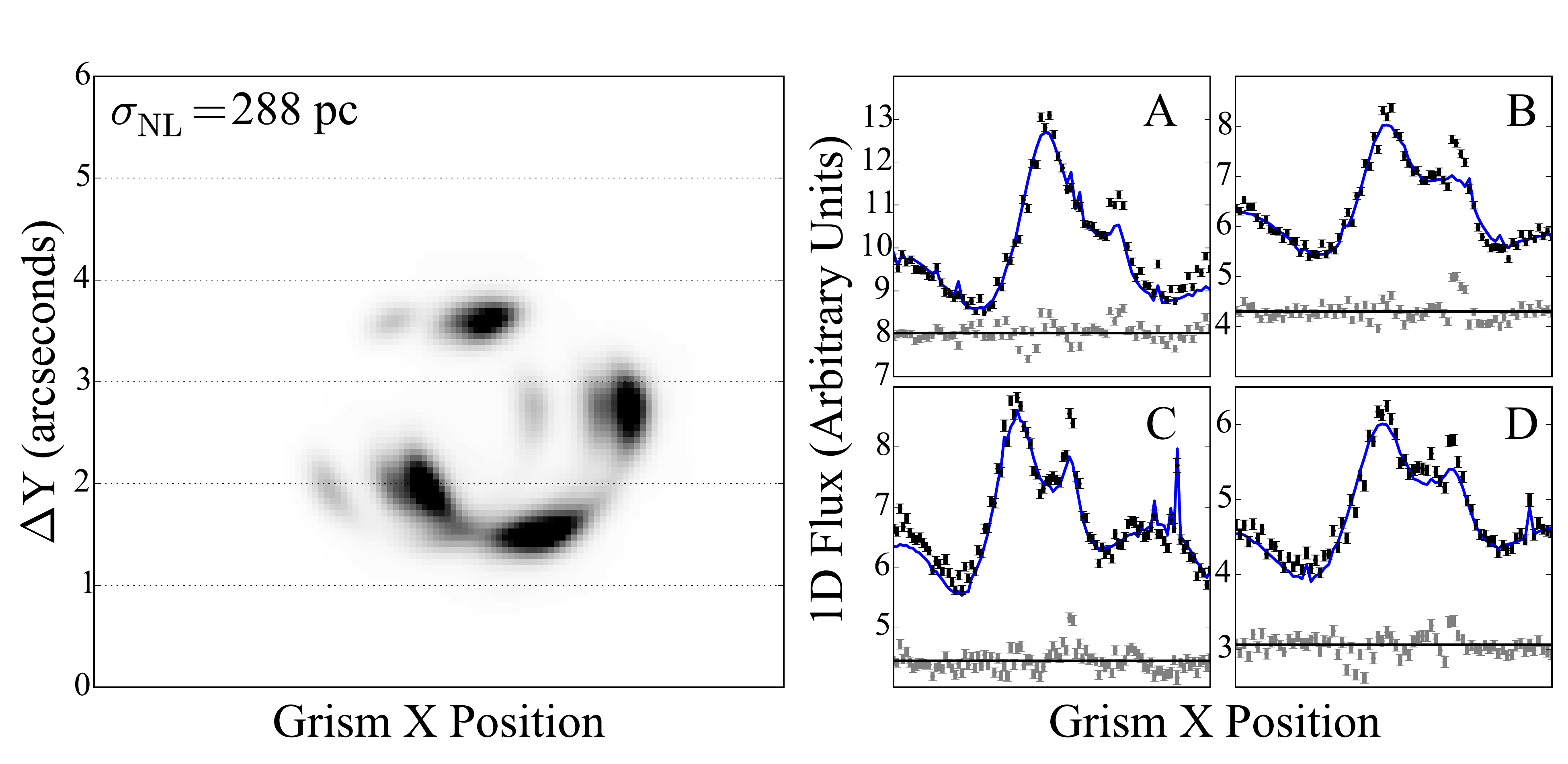} 
\caption{Effects of resolved narrow-line emission on model fit. We repeat the analysis of Section 4, now using a simulated extended narrow-emission region with a Gaussian flux distribution with dispersion $\sigma_{\rm{NL}}=$ 50, 100 and 288 pc. Here we show in the large left panels, the best fit grism model for the extended narrow-line emission with a fixed finite size, after re-optimizing the 1D spectral parameters. While the full spectral model is used in the inference, we show only the narrow-line emission in the left panels for clarity.  In the adjacent right hand panels we show a comparison of the model 1D trace spectra with the data. As in Figure 2, the black points, blue line and light grey points correspond to the data, model and residual respectively. The point source model and comparison between model and data are show in the upper left corner for comparison. The best fit $\chi^2$ values calculated from the difference between the 2D model grism and true grism images are 13228, 13368 and 13990 for 3909 DOF for the 50, 100 and 288 pc narrow-line regions, and 13112 for 3906 DOF for the point source model. The point source model has three extra degrees of freedom as the narrow-line fluxes are allowed to vary independently for each image.}
\label{fig:prmass}
\end{figure*}

We can also test what the effect would be on the inferred flux ratios, if we use a point source model for the narrow-line flux when the flux is actually resolved. For this we considered only the 50 pc and 100 pc models as the 288 pc model provides a markedly worse fit to the data. We simulated a mock spectrum using the resolved narrow-line models for the narrow-line component, and then inferred the image fluxes under the assumption used elsewhere in this work, that the narrow-emission is unresolved in our data. The resulting flux ratios were: A/C $= 0.97\pm 0.08$, B/C $= 0.99\pm$0.07 and D/C $=0.65\pm0.05$ for the 50 pc source, and  
A/C $= 1.0\pm 0.1$, B/C $= 1.0\pm$0.1 and D/C $=0.65\pm0.08$ for the 100 pc source. Both results are consistent with the input fluxes used to generate the resolved narrow-line source mock lenses, indicating that our results would not be biased by a marginally resolved narrow-line region.


\section{Discussion}
We have demonstrated that the WFC3 IR Grism provides sufficient spatial and spectral resolution to precisely measure lensed narrow-line emission with similar precision to continuum emission studies while avoiding the effects of microlensing, variability and differential dust extinction. We measure the [OIII] flux ratios to be significantly offset from optical/near-IR continuum measurements of this system, particularly for image A. This is consistent with the results from long term monitoring for this system, which have indicated that image A is significantly affected by microlensing, which systematically affects smaller emission regions. The [OIII] flux ratios are consistent with the 5 GHz radio measurement from \citet{Jackson++15}, which are also extended and thus not affected by microlensing.

In order to fit the measured F140W image positions and [OIII] fluxes, we rule out the presence of a perturbing NFW subhalo projected within roughly $\sim$1\farcs0 (0\farcs1)  for a $M_{600}\sim 10^8$ $(10^7) M_\odot$ perturber near the lensed images. At the redshift of the lens these angular sizes correspond to approximately $\sim6$ and 0.6 kpc respectively. It is informative to compare these limits with an approximate prediction of the number of subhalos at these radii based on CDM models.

We perform a very basic estimate assuming that all possible perturbers are within the virial radius of the HE0435 group and neglecting line-of-sight associations. From \citet{Sluse++16}, the HE0435 group has a virial mass of $\log[M_{200}/M_\odot] = 13.7\pm0.4$, where $M_{200}$ is the mass within the region of the halo which has a mean density 200 times the critical density. 

We estimated the number of subhalos based on the results of \citet{Han++16}, scaled to the virial mass of HE0435, assuming that approximately half of subhalos are destroyed through tidal interactions and merging following the recommendation of \citet{Han++16}, and that an additional $\sim 30$\% are destroyed by tidal interactions with a central baryonic potential \citep[e.g.][]{Garrison-Kimmel++17}

We assume that the subhalo mass profiles are NFW with the mass-concentration relation given by \citet{Maccio++08} scaled to the redshift of the group following the relation by \citet{Prada++12} and neglecting scatter. We fix $M_{600}$ for subhalos at infall, where $M_{600}$ is the mass within the interior 600 pc of the subhalo. We make a simplifying assumption that $M_{600}$ is not affected by tidal stripping after infall.  This yields an estimated number of $\sim 250$ surviving subhalos in the group with masses greater than $M_{600}>10^8 M_\odot$ and $\sim 19000$ subhalos with  $M_{600}>10^7 M_\odot$ within $R_{200} = 540 $kpc.

We examined two different spatial distributions for the subhalos. First, that the subhalo number density follows the mass distribution of the host halo everywhere, as predicted by pure CDM simulations in the absence of tidal stripping. Second, that the subhalo number density follows the mass distribution of the host halo except within the three-dimensional scale radius, where we assume all subhalos are destroyed. This  mimics an extreme version of the impact of the disk seen by \citet{Garrison-Kimmel++17}. Both spatial distributions are normalized to have the same number of subhalos.  
These two subhalo spatial distributions are chosen to bracket limits of the possible effect of the baryonic potential on our predicted detection rate. 

 In the first case, where the subhalo number density simply follows the NFW profile of the lens halo, we expect approximately $\sim 0.8$ total subhalos with $M_{600}>10^8 M_\odot$ to be found within each $\sim 6$ kpc projected excluded region. 
 The mass function and size of the sensitivity region scales so that we expect to find $\sim1$ $M_{600}>10^7 M_\odot$ subhalo within each of the four smaller $\sim0.6$ kpc projected regions around the images where we are sensitive to these lower masses.
 
In the second case in which all subhalos are removed within the NFW scale radius of the host, we expect to detect approximately 0.08 (0.1) subhalos with $M_{600}>10^8 M_\odot$ ($10^7 M_\odot$) per exclusion region.

We also explored the effect of a WDM mass function, using the result from \citet{Schneider++13} to estimate the shape of the subhalo mass function at infall given a 3 keV/c$^2$ thermal relic dark matter particle, which is consistent with Ly$\alpha$ forest measurements \citep[e.g.][]{Viel++09}. In this case, almost no subhalos with masses  $M_{600}<10^8 M_\odot$ survive. Unlike the central baryonic potential, which is predicted to have a largely mass-independent effect on the subhalo population, WDM selectively destroys low mass subhalos.
Promisingly, subhalos with $M_{600}>10^8 M_\odot$ are all expected to contain a significant number of stars \citep{Strigari++08} in order to match comparisons between the Milky Way satellite luminosity function and CDM predictions \citep[e.g.][]{Strigari++08}.
This mean that CDM can be tested by comparing the rate of detections of substructure in narrow-line lenses with the rate predicted by luminous satellite studies, or measured by gravitational imaging studies which are sensitive to $M_{600}>10^8 M_\odot$ subhalos, after accounting for variations in host halo mass and sensitivity. This test can be performed in a sample of $\sim20$ lenses in which we would expect to detect $\sim10$ $M_{600}>10^7 M_\odot$ subhalos in the case of a CDM subhalo mass function, even in the case in which all subhalos are destroyed within the NFW scale radius.


We emphasise however, that a true comparison requires a CDM model which incorporates effects such as tidal stripping, as well as a marginalization over possible halo orientations, masses and formation histories \citep[e.g.][]{Jiang++16}, and a range of baryonic physics implementations \citep[e.g.][]{Chua++16, Despali++16}. We leave such an analysis to a future paper, in which we jointly infer the properties of the subhalo mass function given our sample of narrow-line gravitational lenses measured with OSIRIS and the WFC3 grism.
 
\section{Summary}
\begin{enumerate}
 \item We present a forward modelling method which uses the {\tt 3D-HST} pipeline \citep{Brammer++12} to measure spectra in the presence of significant spatial blending for G141 grism data. We apply this method to infer the lensed narrow, broad and continuum fluxes of the images in HE0435-1223.
 \item The narrow [OIII] flux ratios for HE0435 are consistent with radio measurements from \citet{Jackson++15}, and are significantly different from other emission measures which are subject to contamination by microlensing and intrinsic QSO time variability.
 \item We find that the [OIII] fluxes and image positions are well modelled with a simple gravitational lens model consisting of a singular isothermal ellipsoid for the main galaxy in the presence of external shear. 
 \item Our data strongly disfavours a perturber with mass greater than $M_{600}=10^{8.2} (10^{7.2}) M_\odot$ within $\sim1$ (0.1) arcsecond of the lensed images, where $M_{600}$ is the projected perturber mass within its central 600 pc (best fit model probability $<0.3$\%).
 
\item This demonstration that WFC3 grism measurement of narrow-line lensed quasars can be used to detect low-mass $M_{600}\sim10^{7}M_\odot$ subhalos is extremely promising for future constraints of dark matter given the large number of quadruply imaged quasar lenses to be discovered in optical surveys such as DES and LSST, and with the follow-up which will be enabled by JWST.

\end{enumerate}

\section*{Acknowledgments}
Based on observations made with the NASA/ESA Hubble Space Telescope, obtained at the Space Telescope Science Institute, which is operated by the Association of Universities for Research in Astronomy, Inc., under NASA contract NAS 5-26555. These observations are associated with program \#13732.
Support for program \#13732 was provided by NASA through a grant from the Space Telescope Science Institute, which is operated by the Association of Universities for Research in Astronomy, Inc., under NASA contract NAS 5-26555.
T.T. thanks the Packard Foundation for generous support through a Packard Research Fellowship. A.M.N thanks the Center of Cosmology and AstroParticle Physics for support via a CCAPP postdoctoral fellowship. C.S.K. is supported by NSF grant AST-1515876.

We thank K. Wong for providing model constraints for the deflector and inferred lensed image fluxes based on F160W data. We thank V. Bonvin and F. Courbin for providing microlensing light curve constraints.

\bibliographystyle{apj_2}
\bibliography{references}

\begin{thebibliography}{103}
\expandafter\ifx\csname natexlab\endcsname\relax\def\natexlab#1{#1}\fi

\bibitem[{{Agnello} {et~al.}(2015){Agnello}, {Treu}, {Ostrovski}, {Schechter},
  {Buckley-Geer}, {Lin}, {Auger}, {Courbin}, {Fassnacht}, {Frieman},
  {Kuropatkin}, {Marshall}, {McMahon}, {Meylan}, {More}, {Suyu}, {Rusu},
  {Finley}, {Abbott}, {Abdalla}, {Allam}, {Annis}, {Banerji},
  {Benoit-L{\'e}vy}, {Bertin}, {Brooks}, {Burke}, {Rosell}, {Kind},
  {Carretero}, {Cunha}, {D'Andrea}, {da Costa}, {Desai}, {Diehl}, {Dietrich},
  {Doel}, {Eifler}, {Estrada}, {Neto}, {Flaugher}, {Fosalba}, {Gerdes},
  {Gruen}, {Gutierrez}, {Honscheid}, {James}, {Kuehn}, {Lahav}, {Lima}, {Maia},
  {March}, {Marshall}, {Martini}, {Melchior}, {Miller}, {Miquel}, {Nichol},
  {Ogando}, {Plazas}, {Reil}, {Romer}, {Roodman}, {Sako}, {Sanchez},
  {Santiago}, {Scarpine}, {Schubnell}, {Sevilla-Noarbe}, {Smith},
  {Soares-Santos}, {Sobreira}, {Suchyta}, {Swanson}, {Tarle}, {Thaler},
  {Tucker}, {Walker}, {Wechsler}, \& {Zhang}}]{Agnello++15}
{Agnello}, A., {Treu}, T., {Ostrovski}, F., {et~al.} 2015, \mnras, 454, 1260

\bibitem[{{Bennert} {et~al.}(2006{\natexlab{a}}){Bennert}, {Jungwiert},
  {Komossa}, {Haas}, \& {Chini}}]{Bennert++06b}
{Bennert}, N., {Jungwiert}, B., {Komossa}, S., {Haas}, M., \& {Chini}, R.
  2006{\natexlab{a}}, \aap, 459, 55

\bibitem[{{Bennert} {et~al.}(2006{\natexlab{b}}){Bennert}, {Jungwiert},
  {Komossa}, {Haas}, \& {Chini}}]{Bennert++06a}
---. 2006{\natexlab{b}}, \aap, 456, 953

\bibitem[{{Bennert} {et~al.}(2011){Bennert}, {Auger}, {Treu}, {Woo}, \&
  {Malkan}}]{Bennert++11}
{Bennert}, V.~N., {Auger}, M.~W., {Treu}, T., {Woo}, J.-H., \& {Malkan}, M.~A.
  2011, \apj, 742, 107

\bibitem[{{Birrer} {et~al.}(2017){Birrer}, {Amara}, \&
  {Refregier}}]{Birrer++17}
{Birrer}, S., {Amara}, A., \& {Refregier}, A. 2017, ArXiv:1702.00009

\bibitem[{{Blackburne} {et~al.}(2014){Blackburne}, {Kochanek}, {Chen}, {Dai},
  \& {Chartas}}]{Blackburne++14}
{Blackburne}, J.~A., {Kochanek}, C.~S., {Chen}, B., {Dai}, X., \& {Chartas}, G.
  2014, \apj, 789, 125

\bibitem[{{Blackburne} {et~al.}(2011){Blackburne}, {Pooley}, {Rappaport}, \&
  {Schechter}}]{Blackburne++11}
{Blackburne}, J.~A., {Pooley}, D., {Rappaport}, S., \& {Schechter}, P.~L. 2011,
  \apj, 729, 34

\bibitem[{{Bonvin} {et~al.}(2017){Bonvin}, {Courbin}, {Suyu}, {Marshall},
  {Rusu}, {Sluse}, {Tewes}, {Wong}, {Collett}, {Fassnacht}, {Treu}, {Auger},
  {Hilbert}, {Koopmans}, {Meylan}, {Rumbaugh}, {Sonnenfeld}, \&
  {Spiniello}}]{Bonvin++16}
{Bonvin}, V., {Courbin}, F., {Suyu}, S.~H., {et~al.} 2017, \mnras, 465, 4914

\bibitem[{{Brammer} {et~al.}(2012){Brammer}, {van Dokkum}, {Franx},
  {Fumagalli}, {Patel}, {Rix}, {Skelton}, {Kriek}, {Nelson}, {Schmidt},
  {Bezanson}, {da Cunha}, {Erb}, {Fan}, {F{\"o}rster Schreiber}, {Illingworth},
  {Labb{\'e}}, {Leja}, {Lundgren}, {Magee}, {Marchesini}, {McCarthy},
  {Momcheva}, {Muzzin}, {Quadri}, {Steidel}, {Tal}, {Wake}, {Whitaker}, \&
  {Williams}}]{Brammer++12}
{Brammer}, G.~B., {van Dokkum}, P.~G., {Franx}, M., {et~al.} 2012, \apjs, 200,
  13

\bibitem[{{Brooks} {et~al.}(2013){Brooks}, {Kuhlen}, {Zolotov}, \&
  {Hooper}}]{Brooks++13}
{Brooks}, A.~M., {Kuhlen}, M., {Zolotov}, A., \& {Hooper}, D. 2013, \apj, 765,
  22

\bibitem[{{Chiba} {et~al.}(2005){Chiba}, {Minezaki}, {Kashikawa}, {Kataza}, \&
  {Inoue}}]{Chiba++05}
{Chiba}, M., {Minezaki}, T., {Kashikawa}, N., {Kataza}, H., \& {Inoue}, K.~T.
  2005, \apj, 627, 53

\bibitem[{{Chua} {et~al.}(2016){Chua}, {Pillepich}, {Rodriguez-Gomez},
  {Vogelsberger}, {Bird}, \& {Hernquist}}]{Chua++16}
{Chua}, K.~T.~E., {Pillepich}, A., {Rodriguez-Gomez}, V., {et~al.} 2016,
  ArXiv:1611.07991

\bibitem[{{Courbin} {et~al.}(2011){Courbin}, {Chantry}, {Revaz}, {Sluse},
  {Faure}, {Tewes}, {Eulaers}, {Koleva}, {Asfandiyarov}, {Dye}, {Magain}, {van
  Winckel}, {Coles}, {Saha}, {Ibrahimov}, \& {Meylan}}]{Courbin++11}
{Courbin}, F., {Chantry}, V., {Revaz}, Y., {et~al.} 2011, \aap, 536, A53

\bibitem[{{Dalal} \& {Kochanek}(2002)}]{Dalal++02}
{Dalal}, N., \& {Kochanek}, C.~S. 2002, \apj, 572, 25

\bibitem[{{Despali} \& {Vegetti}(2016)}]{Despali++16}
{Despali}, G., \& {Vegetti}, S. 2016, ArXiv:1608.06938

\bibitem[{{Diemand} {et~al.}(2008){Diemand}, {Kuhlen}, {Madau}, {Zemp},
  {Moore}, {Potter}, \& {Stadel}}]{Diemand++08}
{Diemand}, J., {Kuhlen}, M., {Madau}, P., {et~al.} 2008, \nat, 454, 735

\bibitem[{{Drlica-Wagner} {et~al.}(2015){Drlica-Wagner}, {Bechtol}, {Rykoff},
  {Luque}, {Queiroz}, {Mao}, {Wechsler}, {Simon}, {Santiago}, {Yanny},
  {Balbinot}, {Dodelson}, {Fausti Neto}, {James}, {Li}, {Maia}, {Marshall},
  {Pieres}, {Stringer}, {Walker}, {Abbott}, {Abdalla}, {Allam},
  {Benoit-L{\'e}vy}, {Bernstein}, {Bertin}, {Brooks}, {Buckley-Geer}, {Burke},
  {Carnero Rosell}, {Carrasco Kind}, {Carretero}, {Crocce}, {da Costa},
  {Desai}, {Diehl}, {Dietrich}, {Doel}, {Eifler}, {Evrard}, {Finley},
  {Flaugher}, {Fosalba}, {Frieman}, {Gaztanaga}, {Gerdes}, {Gruen}, {Gruendl},
  {Gutierrez}, {Honscheid}, {Kuehn}, {Kuropatkin}, {Lahav}, {Martini},
  {Miquel}, {Nord}, {Ogando}, {Plazas}, {Reil}, {Roodman}, {Sako}, {Sanchez},
  {Scarpine}, {Schubnell}, {Sevilla-Noarbe}, {Smith}, {Soares-Santos},
  {Sobreira}, {Suchyta}, {Swanson}, {Tarle}, {Tucker}, {Vikram}, {Wester},
  {Zhang}, {Zuntz}, \& {DES Collaboration}}]{Drlica-Wagner++15}
{Drlica-Wagner}, A., {Bechtol}, K., {Rykoff}, E.~S., {et~al.} 2015, \apj, 813,
  109

\bibitem[{{Dunkley} {et~al.}(2009){Dunkley}, {Spergel}, {Komatsu}, {Hinshaw},
  {Larson}, {Nolta}, {Odegard}, {Page}, {Bennett}, {Gold}, {Hill}, {Jarosik},
  {Weiland}, {Halpern}, {Kogut}, {Limon}, {Meyer}, {Tucker}, {Wollack}, \&
  {Wright}}]{Dunkley++09}
{Dunkley}, J., {Spergel}, D.~N., {Komatsu}, E., {et~al.} 2009, \apj, 701, 1804

\bibitem[{{Dutton} {et~al.}(2011){Dutton}, {Brewer}, {Marshall}, {Auger},
  {Treu}, {Koo}, {Bolton}, {Holden}, \& {Koopmans}}]{Dutton++11}
{Dutton}, A.~A., {Brewer}, B.~J., {Marshall}, P.~J., {et~al.} 2011, \mnras,
  417, 1621

\bibitem[{{Fadely} \& {Keeton}(2011)}]{Fadely++11}
{Fadely}, R., \& {Keeton}, C.~R. 2011, \aj, 141, 101

\bibitem[{{Fadely} \& {Keeton}(2012)}]{Fadely++12}
---. 2012, \mnras, 419, 936

\bibitem[{{Falco} {et~al.}(1999){Falco}, {Impey}, {Kochanek}, {Leh{\'a}r},
  {McLeod}, {Rix}, {Keeton}, {Mu{\~n}oz}, \& {Peng}}]{Fal++99}
{Falco}, E.~E., {Impey}, C.~D., {Kochanek}, C.~S., {et~al.} 1999, \apj, 523,
  617

\bibitem[{{Ferrari} {et~al.}(1999){Ferrari}, {Pastoriza}, {Macchetto}, \&
  {Caon}}]{Ferrari++99}
{Ferrari}, F., {Pastoriza}, M.~G., {Macchetto}, F., \& {Caon}, N. 1999, \aaps,
  136, 269

\bibitem[{{Foreman-Mackey} {et~al.}(2013){Foreman-Mackey}, {Hogg}, {Lang}, \&
  {Goodman}}]{Foreman-Mackey++13}
{Foreman-Mackey}, D., {Hogg}, D.~W., {Lang}, D., \& {Goodman}, J. 2013, \pasp,
  125, 306

\bibitem[{{Garrison-Kimmel} {et~al.}(2017){Garrison-Kimmel}, {Wetzel},
  {Bullock}, {Hopkins}, {Boylan-Kolchin}, {Faucher-Giguere}, {Keres},
  {Quataert}, {Sanderson}, {Graus}, \& {Kelley}}]{Garrison-Kimmel++17}
{Garrison-Kimmel}, S., {Wetzel}, A.~R., {Bullock}, J.~S., {et~al.} 2017,
  ArXiv:1701.03792

\bibitem[{{Gavazzi} {et~al.}(2007){Gavazzi}, {Treu}, {Rhodes}, {Koopmans},
  {Bolton}, {Burles}, {Massey}, \& {Moustakas}}]{Gavazzi++07}
{Gavazzi}, R., {Treu}, T., {Rhodes}, J.~D., {et~al.} 2007, \apj, 667, 176

\bibitem[{{Gilman} {et~al.}(2017){Gilman}, {Agnello}, {Treu}, {Keeton}, \&
  {Nierenberg}}]{Gilman++16}
{Gilman}, D., {Agnello}, A., {Treu}, T., {Keeton}, C.~R., \& {Nierenberg},
  A.~M. 2017, \mnras, 467, 3970

\bibitem[{{Gnedin}(2000)}]{Gnedin++00}
{Gnedin}, N.~Y. 2000, \apjl, 535, L75

\bibitem[{{Gonzaga} \& {et al.}(2012)}]{Gonzaga++12}
{Gonzaga}, S., \& {et al.} 2012, {The DrizzlePac Handbook}

\bibitem[{{Guo} {et~al.}(2011){Guo}, {Cole}, {Eke}, \& {Frenk}}]{Guo++11}
{Guo}, Q., {Cole}, S., {Eke}, V., \& {Frenk}, C. 2011, \mnras, 1278

\bibitem[{{Han} {et~al.}(2016){Han}, {Cole}, {Frenk}, \& {Jing}}]{Han++16}
{Han}, J., {Cole}, S., {Frenk}, C.~S., \& {Jing}, Y. 2016, \mnras, 457, 1208

\bibitem[{{Hargis} {et~al.}(2014){Hargis}, {Willman}, \& {Peter}}]{Hargis++14}
{Hargis}, J.~R., {Willman}, B., \& {Peter}, A.~H.~G. 2014, \apjl, 795, L13

\bibitem[{{Hezaveh} {et~al.}(2016){Hezaveh}, {Dalal}, {Marrone}, {Mao},
  {Morningstar}, {Wen}, {Blandford}, {Carlstrom}, {Fassnacht}, {Holder},
  {Kemball}, {Marshall}, {Murray}, {Perreault Levasseur}, {Vieira}, \&
  {Wechsler}}]{Hezaveh++16}
{Hezaveh}, Y.~D., {Dalal}, N., {Marrone}, D.~P., {et~al.} 2016, \apj, 823, 37

\bibitem[{{H{\"o}nig} {et~al.}(2008){H{\"o}nig}, {Smette}, {Beckert}, {Horst},
  {Duschl}, {Gandhi}, {Kishimoto}, \& {Weigelt}}]{Honig++08}
{H{\"o}nig}, S.~F., {Smette}, A., {Beckert}, T., {et~al.} 2008, \aap, 485, L21

\bibitem[{{Hsueh} {et~al.}(2016){Hsueh}, {Fassnacht}, {Vegetti}, {McKean},
  {Spingola}, {Auger}, {Koopmans}, \& {Lagattuta}}]{Hsueh++16}
{Hsueh}, J.-W., {Fassnacht}, C.~D., {Vegetti}, S., {et~al.} 2016, \mnras, 463,
  L51

\bibitem[{{Jackson} {et~al.}(2015){Jackson}, {Tagore}, {Roberts}, {Sluse},
  {Stacey}, {Vives-Arias}, {Wucknitz}, \& {Volino}}]{Jackson++15}
{Jackson}, N., {Tagore}, A.~S., {Roberts}, C., {et~al.} 2015, \mnras, 454, 287

\bibitem[{{Jiang} \& {van den Bosch}(2016)}]{Jiang++16}
{Jiang}, F., \& {van den Bosch}, F.~C. 2016, ArXiv:1610.02399

\bibitem[{{Kaufmann} {et~al.}(2008){Kaufmann}, {Bullock}, {Maller}, \&
  {Fang}}]{Kaufmann++08}
{Kaufmann}, T., {Bullock}, J.~S., {Maller}, A., \& {Fang}, T. 2008, in American
  Institute of Physics Conference Series, Vol. 1035, The Evolution of Galaxies
  Through the Neutral Hydrogen Window, ed. {R.~Minchin \& E.~Momjian}, 147--150

\bibitem[{{Keeton}(2001{\natexlab{a}})}]{Keeton++01b}
{Keeton}, C.~R. 2001{\natexlab{a}}, ArXiv:astro-ph/0102341

\bibitem[{{Keeton}(2001{\natexlab{b}})}]{Keeton++01a}
---. 2001{\natexlab{b}}, ArXiv:astro-ph/0102340

\bibitem[{{Keeton}(2009)}]{Keeton++09a}
---. 2009, ArXiv:0908.3001

\bibitem[{{Keeton} \& {Moustakas}(2009)}]{Keeton++09b}
{Keeton}, C.~R., \& {Moustakas}, L.~A. 2009, \apj, 699, 1720

\bibitem[{{Klypin} {et~al.}(1999){Klypin}, {Kravtsov}, {Valenzuela}, \&
  {Prada}}]{Klypin++99}
{Klypin}, A., {Kravtsov}, A.~V., {Valenzuela}, O., \& {Prada}, F. 1999, \apj,
  522, 82

\bibitem[{{Kochanek} {et~al.}(2006){Kochanek}, {Morgan}, {Falco}, {McLeod},
  {Winn}, {Dembicky}, \& {Ketzeback}}]{Kochanek++06}
{Kochanek}, C.~S., {Morgan}, N.~D., {Falco}, E.~E., {et~al.} 2006, \apj, 640,
  47

\bibitem[{{Koekemoer} {et~al.}(2011){Koekemoer}, {Faber}, {Ferguson}, {Grogin},
  {Kocevski}, {Koo}, {Lai}, {Lotz}, {Lucas}, {McGrath}, {Ogaz}, {Rajan},
  {Riess}, {Rodney}, {Strolger}, {Casertano}, {Castellano}, {Dahlen}, \&
  {et}}]{Koekemoer++11}
{Koekemoer}, A.~M., {Faber}, S.~M., {Ferguson}, H.~C., {et~al.} 2011, \apjs,
  197, 36

\bibitem[{{Larkin} {et~al.}(2006){Larkin}, {Barczys}, {Krabbe}, {Adkins},
  {Aliado}, {Amico}, {Brims}, {Campbell}, {Canfield}, {Gasaway}, {Honey},
  {Iserlohe}, {Johnson}, {Kress}, {LaFreniere}, {Lyke}, {Magnone}, {Magnone},
  {McElwain}, {Moon}, {Quirrenbach}, {Skulason}, {Song}, {Spencer}, {Weiss}, \&
  {Wright}}]{Larkin++06}
{Larkin}, J., {Barczys}, M., {Krabbe}, A., {et~al.} 2006, in ProcSPIE, Vol.
  6269, Society of Photo-Optical Instrumentation Engineers (SPIE) Conference
  Series, 62691A

\bibitem[{{Lu} {et~al.}(2014){Lu}, {Wechsler}, {Somerville}, {Croton},
  {Porter}, {Primack}, {Behroozi}, {Ferguson}, {Koo}, {Guo}, {Safarzadeh},
  {Finlator}, {Castellano}, {White}, {Sommariva}, \& {Moody}}]{Lu++14}
{Lu}, Y., {Wechsler}, R.~H., {Somerville}, R.~S., {et~al.} 2014, \apj, 795, 123

\bibitem[{{Macci{\`o}} {et~al.}(2008){Macci{\`o}}, {Dutton}, \& {van den
  Bosch}}]{Maccio++08}
{Macci{\`o}}, A.~V., {Dutton}, A.~A., \& {van den Bosch}, F.~C. 2008, \mnras,
  391, 1940

\bibitem[{{Macci{\`o}} {et~al.}(2010){Macci{\`o}}, {Kang}, {Fontanot},
  {Somerville}, {Koposov}, \& {Monaco}}]{Maccio++11}
{Macci{\`o}}, A.~V., {Kang}, X., {Fontanot}, F., {et~al.} 2010, \mnras, 402,
  1995

\bibitem[{{MacLeod} {et~al.}(2009){MacLeod}, {Kochanek}, \&
  {Agol}}]{Macleod++09}
{MacLeod}, C.~L., {Kochanek}, C.~S., \& {Agol}, E. 2009, \apj, 699, 1578

\bibitem[{{Menci} {et~al.}(2014){Menci}, {Gatti}, {Fiore}, \&
  {Lamastra}}]{Menci++14}
{Menci}, N., {Gatti}, M., {Fiore}, F., \& {Lamastra}, A. 2014, \aap, 569, A37

\bibitem[{{Minezaki} {et~al.}(2009){Minezaki}, {Chiba}, {Kashikawa}, {Inoue},
  \& {Kataza}}]{Minezaki++09}
{Minezaki}, T., {Chiba}, M., {Kashikawa}, N., {Inoue}, K.~T., \& {Kataza}, H.
  2009, \apj, 697, 610

\bibitem[{{Momcheva} {et~al.}(2006){Momcheva}, {Williams}, {Keeton}, \&
  {Zabludoff}}]{Momcheva++06}
{Momcheva}, I., {Williams}, K., {Keeton}, C., \& {Zabludoff}, A. 2006, \apj,
  641, 169

\bibitem[{{Momcheva} {et~al.}(2015){Momcheva}, {Williams}, {Cool}, {Keeton}, \&
  {Zabludoff}}]{Momcheva++15}
{Momcheva}, I.~G., {Williams}, K.~A., {Cool}, R.~J., {Keeton}, C.~R., \&
  {Zabludoff}, A.~I. 2015, \apjs, 219, 29

\bibitem[{{Momcheva} {et~al.}(2016){Momcheva}, {Brammer}, {van Dokkum},
  {Skelton}, {Whitaker}, {Nelson}, {Fumagalli}, {Maseda}, {Leja}, {Franx},
  {Rix}, {Bezanson}, {Da Cunha}, {Dickey}, {F{\"o}rster Schreiber},
  {Illingworth}, {Kriek}, {Labb{\'e}}, {Ulf Lange}, {Lundgren}, {Magee},
  {Marchesini}, {Oesch}, {Pacifici}, {Patel}, {Price}, {Tal}, {Wake}, {van der
  Wel}, \& {Wuyts}}]{Momcheva++16}
{Momcheva}, I.~G., {Brammer}, G.~B., {van Dokkum}, P.~G., {et~al.} 2016, \apjs,
  225, 27

\bibitem[{{Moore} {et~al.}(1999){Moore}, {Ghigna}, {Governato}, {Lake},
  {Quinn}, {Stadel}, \& {Tozzi}}]{Moore++1999}
{Moore}, B., {Ghigna}, S., {Governato}, F., {et~al.} 1999, \apjl, 524, L19

\bibitem[{{Morgan} {et~al.}(2005){Morgan}, {Kochanek}, {Pevunova}, \&
  {Schechter}}]{Morgan++05}
{Morgan}, N.~D., {Kochanek}, C.~S., {Pevunova}, O., \& {Schechter}, P.~L. 2005,
  \aj, 129, 2531

\bibitem[{{Moustakas} \& {Metcalf}(2003)}]{Moustakas++03}
{Moustakas}, L.~A., \& {Metcalf}, R.~B. 2003, \mnras, 339, 607

\bibitem[{{M{\"u}ller-S{\'a}nchez} {et~al.}(2011){M{\"u}ller-S{\'a}nchez},
  {Prieto}, {Hicks}, {Vives-Arias}, {Davies}, {Malkan}, {Tacconi}, \&
  {Genzel}}]{Muller-Sanchez++11}
{M{\"u}ller-S{\'a}nchez}, F., {Prieto}, M.~A., {Hicks}, E.~K.~S., {et~al.}
  2011, \apj, 739, 69

\bibitem[{{Navarro} {et~al.}(1996){Navarro}, {Frenk}, \& {White}}]{NFW++1996}
{Navarro}, J.~F., {Frenk}, C.~S., \& {White}, S.~D.~M. 1996, \apj, 462, 563

\bibitem[{{Nierenberg} {et~al.}(2016){Nierenberg}, {Treu}, {Menci}, {Lu},
  {Torrey}, \& {Vogelsberger}}]{Nierenberg++16}
{Nierenberg}, A.~M., {Treu}, T., {Menci}, N., {et~al.} 2016, \mnras, 462, 4473

\bibitem[{{Nierenberg} {et~al.}(2013){Nierenberg}, {Treu}, {Menci}, {Lu}, \&
  {Wang}}]{Nierenberg++13}
{Nierenberg}, A.~M., {Treu}, T., {Menci}, N., {Lu}, Y., \& {Wang}, W. 2013,
  \apj, 772, 146

\bibitem[{{Nierenberg} {et~al.}(2014){Nierenberg}, {Treu}, {Wright},
  {Fassnacht}, \& {Auger}}]{Nierenberg++14}
{Nierenberg}, A.~M., {Treu}, T., {Wright}, S.~A., {Fassnacht}, C.~D., \&
  {Auger}, M.~W. 2014, \mnras, 442, 2434

\bibitem[{Oguri \& Marshall(2010)}]{Oguri++10}
Oguri, M., \& Marshall, P.~J. 2010, \mnras, 405, 2579

\bibitem[{{Oke}(1974)}]{Oke++1974}
{Oke}, J.~B. 1974, \apjs, 27, 21

\bibitem[{{Ostrovski} {et~al.}(2017){Ostrovski}, {McMahon}, {Connolly},
  {Lemon}, {Auger}, {Banerji}, {Hung}, {Koposov}, {Lidman}, {Reed}, {Allam},
  {Benoit-L{\'e}vy}, {Bertin}, {Brooks}, {Buckley-Geer}, {Carnero Rosell},
  {Carrasco Kind}, {Carretero}, {Cunha}, {da Costa}, {Desai}, {Diehl},
  {Dietrich}, {Evrard}, {Finley}, {Flaugher}, {Fosalba}, {Frieman}, {Gerdes},
  {Goldstein}, {Gruen}, {Gruendl}, {Gutierrez}, {Honscheid}, {James}, {Kuehn},
  {Kuropatkin}, {Lima}, {Lin}, {Maia}, {Marshall}, {Martini}, {Melchior},
  {Miquel}, {Ogando}, {Plazas Malag{\'o}n}, {Reil}, {Romer}, {Sanchez},
  {Santiago}, {Scarpine}, {Sevilla-Noarbe}, {Soares-Santos}, {Sobreira},
  {Suchyta}, {Tarle}, {Thomas}, {Tucker}, \& {Walker}}]{Ostrovski++16}
{Ostrovski}, F., {McMahon}, R.~G., {Connolly}, A.~J., {et~al.} 2017, \mnras,
  465, 4325

\bibitem[{{Patnaik} {et~al.}(1992){Patnaik}, {Browne}, {Walsh}, {Chaffee}, \&
  {Foltz}}]{Patnaik++92}
{Patnaik}, A.~R., {Browne}, I.~W.~A., {Walsh}, D., {Chaffee}, F.~H., \&
  {Foltz}, C.~B. 1992, \mnras, 259, 1P

\bibitem[{{Peng} {et~al.}(2002){Peng}, {Ho}, {Impey}, \& {Rix}}]{Peng++02}
{Peng}, C.~Y., {Ho}, L.~C., {Impey}, C.~D., \& {Rix}, H.-W. 2002, \aj, 124, 266

\bibitem[{{Peng} {et~al.}(2010){Peng}, {Ho}, {Impey}, \& {Rix}}]{Peng++10}
---. 2010, \aj, 139, 2097

\bibitem[{{Planck Collaboration} {et~al.}(2014){Planck Collaboration}, {Ade},
  {Aghanim}, {Armitage-Caplan}, {Arnaud}, {Ashdown}, {Atrio-Barandela},
  {Aumont}, {Baccigalupi}, {Banday}, \& et~al.}]{Planck++13}
{Planck Collaboration}, {Ade}, P.~A.~R., {Aghanim}, N., {et~al.} 2014, \aap,
  571, A16

\bibitem[{{Prada} {et~al.}(2012){Prada}, {Klypin}, {Cuesta}, {Betancort-Rijo},
  \& {Primack}}]{Prada++12}
{Prada}, F., {Klypin}, A.~A., {Cuesta}, A.~J., {Betancort-Rijo}, J.~E., \&
  {Primack}, J. 2012, \mnras, 423, 3018

\bibitem[{{Ricci} {et~al.}(2011){Ricci}, {Poels}, {Elyiv}, {Finet}, {Sprimont},
  {Anguita}, {Bozza}, {Browne}, {Burgdorf}, {Calchi Novati}, {Dominik},
  {Dreizler}, {Glitrup}, {Grundahl}, {Harps{\o}e}, {Hessman}, {Hinse},
  {Hornstrup}, {Hundertmark}, {J{\o}rgensen}, {Liebig}, {Maier}, {Mancini},
  {Masi}, {Mathiasen}, {Rahvar}, {Scarpetta}, {Skottfelt}, {Snodgrass},
  {Southworth}, {Teuber}, {Th{\"o}ne}, {Wambsgan{\ss}}, {Zimmer}, {Zub}, \&
  {Surdej}}]{Ricci++11}
{Ricci}, D., {Poels}, J., {Elyiv}, A., {et~al.} 2011, \aap, 528, A42

\bibitem[{{Rusin} \& {Kochanek}(2005)}]{Rusin++05}
{Rusin}, D., \& {Kochanek}, C.~S. 2005, \apj, 623, 666

\bibitem[{{Rusin} {et~al.}(2003){Rusin}, {Kochanek}, \& {Keeton}}]{Rusin++03}
{Rusin}, D., {Kochanek}, C.~S., \& {Keeton}, C.~R. 2003, \apj, 595, 29

\bibitem[{{Schmidt} {et~al.}(2014){Schmidt}, {Treu}, {Brammer}, {Brada{\v c}},
  {Wang}, {Dijkstra}, {Dressler}, {Fontana}, {Gavazzi}, {Henry}, {Hoag},
  {Jones}, {Kelly}, {Malkan}, {Mason}, {Pentericci}, {Poggianti}, {Stiavelli},
  {Trenti}, {von der Linden}, \& {Vulcani}}]{Schmidt++14}
{Schmidt}, K.~B., {Treu}, T., {Brammer}, G.~B., {et~al.} 2014, \apjl, 782, L36

\bibitem[{{Schneider} {et~al.}(2013){Schneider}, {Smith}, \&
  {Reed}}]{Schneider++13}
{Schneider}, A., {Smith}, R.~E., \& {Reed}, D. 2013, \mnras, 433, 1573

\bibitem[{{Sluse} {et~al.}(2007){Sluse}, {Claeskens}, {Hutsem{\'e}kers}, \&
  {Surdej}}]{Sluse++07}
{Sluse}, D., {Claeskens}, J.-F., {Hutsem{\'e}kers}, D., \& {Surdej}, J. 2007,
  \aap, 468, 885

\bibitem[{{Sluse} {et~al.}(2012){Sluse}, {Hutsem{\'e}kers}, {Courbin},
  {Meylan}, \& {Wambsganss}}]{Sluse++12}
{Sluse}, D., {Hutsem{\'e}kers}, D., {Courbin}, F., {Meylan}, G., \&
  {Wambsganss}, J. 2012, \aap, 544, A62

\bibitem[{{Sluse} {et~al.}(2013){Sluse}, {Kishimoto}, {Anguita}, {Wucknitz}, \&
  {Wambsganss}}]{Sluse++13}
{Sluse}, D., {Kishimoto}, M., {Anguita}, T., {Wucknitz}, O., \& {Wambsganss},
  J. 2013, \aap, 553, A53

\bibitem[{{Sluse} {et~al.}(2016){Sluse}, {Sonnenfeld}, {Rumbaugh}, {Rusu},
  {Fassnacht}, {Treu}, {Suyu}, {Wong}, {Auger}, {Bonvin}, {Collett}, {Courbin},
  {Hilbert}, {Koopmans}, {Marshall}, {Meylan}, {Spiniello}, \&
  {Tewes}}]{Sluse++16}
{Sluse}, D., {Sonnenfeld}, A., {Rumbaugh}, N., {et~al.} 2016, ArXiv:1607.00382

\bibitem[{{Springel}(2010)}]{Springel++10}
{Springel}, V. 2010, \araa, 48, 391

\bibitem[{{Springel} {et~al.}(2008){Springel}, {Wang}, {Vogelsberger},
  {Ludlow}, {Jenkins}, {Helmi}, {Navarro}, {Frenk}, \& {White}}]{Springel++08}
{Springel}, V., {Wang}, J., {Vogelsberger}, M., {et~al.} 2008, \mnras, 391,
  1685

\bibitem[{{Starkenburg} {et~al.}(2013){Starkenburg}, {Helmi}, {De Lucia}, {Li},
  {Navarro}, {Font}, {Frenk}, {Springel}, {Vera-Ciro}, \&
  {White}}]{Starkenburg++13}
{Starkenburg}, E., {Helmi}, A., {De Lucia}, G., {et~al.} 2013, \mnras, 429, 725

\bibitem[{{Strigari} {et~al.}(2007){Strigari}, {Bullock}, {Kaplinghat},
  {Diemand}, {Kuhlen}, \& {Madau}}]{Strigari++07}
{Strigari}, L.~E., {Bullock}, J.~S., {Kaplinghat}, M., {et~al.} 2007, \apj,
  669, 676

\bibitem[{{Strigari} {et~al.}(2008){Strigari}, {Bullock}, {Kaplinghat},
  {Simon}, {Geha}, {Willman}, \& {Walker}}]{Strigari++08}
---. 2008, \nat, 454, 1096

\bibitem[{{Sugai} {et~al.}(2007){Sugai}, {Kawai}, {Shimono}, {Hattori},
  {Kosugi}, {Kashikawa}, {Inoue}, \& {Chiba}}]{Sugai++07}
{Sugai}, H., {Kawai}, A., {Shimono}, A., {et~al.} 2007, \apj, 660, 1016

\bibitem[{{Thoul} \& {Weinberg}(1996)}]{Thoul++1996}
{Thoul}, A.~A., \& {Weinberg}, D.~H. 1996, \apj, 465, 608

\bibitem[{{Treu}(2010)}]{Treu++10}
{Treu}, T. 2010, \araa, 48, 87

\bibitem[{{Vegetti} {et~al.}(2014){Vegetti}, {Koopmans}, {Auger}, {Treu}, \&
  {Bolton}}]{Vegetti++14}
{Vegetti}, S., {Koopmans}, L.~V.~E., {Auger}, M.~W., {Treu}, T., \& {Bolton},
  A.~S. 2014, \mnras, 442, 2017

\bibitem[{{Vegetti} {et~al.}(2012){Vegetti}, {Lagattuta}, {McKean}, {Auger},
  {Fassnacht}, \& {Koopmans}}]{Vegetti++12}
{Vegetti}, S., {Lagattuta}, D.~J., {McKean}, J.~P., {et~al.} 2012, \nat, 481,
  341

\bibitem[{{Viel} {et~al.}(2009){Viel}, {Bolton}, \& {Haehnelt}}]{Viel++09}
{Viel}, M., {Bolton}, J.~S., \& {Haehnelt}, M.~G. 2009, \mnras, 399, L39

\bibitem[{{Weinberg} {et~al.}(2008){Weinberg}, {Colombi}, {Dav{\'e}}, \&
  {Katz}}]{Weinberg++08}
{Weinberg}, D.~H., {Colombi}, S., {Dav{\'e}}, R., \& {Katz}, N. 2008, \apj,
  678, 6

\bibitem[{{Wetzel} {et~al.}(2016){Wetzel}, {Hopkins}, {Kim},
  {Faucher-Gigu{\`e}re}, {Kere{\v s}}, \& {Quataert}}]{Wetzel++16}
{Wetzel}, A.~R., {Hopkins}, P.~F., {Kim}, J.-h., {et~al.} 2016, \apjl, 827, L23

\bibitem[{{Weymann} {et~al.}(1980){Weymann}, {Latham}, {Roger}, {Angel},
  {Green}, {Liebert}, {Turnshek}, {Turnshek}, \& {Tyson}}]{Weyman++80}
{Weymann}, R.~J., {Latham}, D., {Roger}, J., {et~al.} 1980, \nat, 285, 641

\bibitem[{{Wilson} {et~al.}(2016){Wilson}, {Zabludoff}, {Ammons}, {Momcheva},
  {Williams}, \& {Keeton}}]{Wilson++16}
{Wilson}, M.~L., {Zabludoff}, A.~I., {Ammons}, S.~M., {et~al.} 2016, \apj, 833,
  194

\bibitem[{{Wisotzki} {et~al.}(2003){Wisotzki}, {Becker}, {Christensen},
  {Helms}, {Jahnke}, {Kelz}, {Roth}, \& {Sanchez}}]{Wisotzki++03}
{Wisotzki}, L., {Becker}, T., {Christensen}, L., {et~al.} 2003, \aap, 408, 455

\bibitem[{{Wisotzki} {et~al.}(2002){Wisotzki}, {Schechter}, {Bradt},
  {Heinm{\"u}ller}, \& {Reimers}}]{Wisotzki++02}
{Wisotzki}, L., {Schechter}, P.~L., {Bradt}, H.~V., {Heinm{\"u}ller}, J., \&
  {Reimers}, D. 2002, \aap, 395, 17

\bibitem[{{Wittkowski} {et~al.}(2004){Wittkowski}, {Kervella}, {Arsenault},
  {Paresce}, {Beckert}, \& {Weigelt}}]{Wittkowski++04}
{Wittkowski}, M., {Kervella}, P., {Arsenault}, R., {et~al.} 2004, \aap, 418,
  L39

\bibitem[{{Wong} {et~al.}(2011){Wong}, {Keeton}, {Williams}, {Momcheva}, \&
  {Zabludoff}}]{Wong++11}
{Wong}, K.~C., {Keeton}, C.~R., {Williams}, K.~A., {Momcheva}, I.~G., \&
  {Zabludoff}, A.~I. 2011, \apj, 726, 84

\bibitem[{{Wong} {et~al.}(2017){Wong}, {Suyu}, {Auger}, {Bonvin}, {Courbin},
  {Fassnacht}, {Halkola}, {Rusu}, {Sluse}, {Sonnenfeld}, {Treu}, {Collett},
  {Hilbert}, {Koopmans}, {Marshall}, \& {Rumbaugh}}]{Wong++16}
{Wong}, K.~C., {Suyu}, S.~H., {Auger}, M.~W., {et~al.} 2017, \mnras, 465, 4895

\bibitem[{{Woo} {et~al.}(2006){Woo}, {Treu}, {Malkan}, \&
  {Blandford}}]{Woo++06}
{Woo}, J.-H., {Treu}, T., {Malkan}, M.~A., \& {Blandford}, R.~D. 2006, \apj,
  645, 900

\bibitem[{{Xu} {et~al.}(2015){Xu}, {Sluse}, {Gao}, {Wang}, {Frenk}, {Mao},
  {Schneider}, \& {Springel}}]{Xu++15}
{Xu}, D., {Sluse}, D., {Gao}, L., {et~al.} 2015, \mnras, 447, 3189

\bibitem[{{Zolotov} {et~al.}(2012){Zolotov}, {Brooks}, {Willman}, {Governato},
  {Pontzen}, {Christensen}, {Dekel}, {Quinn}, {Shen}, \&
  {Wadsley}}]{Zolotov++12}
{Zolotov}, A., {Brooks}, A.~M., {Willman}, B., {et~al.} 2012, \apj, 761, 71

\end{thebibliography}

\appendix
\onecolumn
\section{Table of measurements of flux ratios for HE0435-1223}
Here we provide the F140W lensed quasar image position as well as a selection of the many measurements of system chosen to illustrate how the flux ratios vary with wavelength and time, as shown in Figures 3 and 4.

\begin{table}
\centering
\scriptsize\begin{tabular}{llllll}
\hline
Image  &  dRa & dDec & Uncertainty \\
\hline
A           & 2.476 & 0.608 & 0.008 \\
B           & 0.997 & 1.157 & 0.008 \\
C           & 0        & 0        & 0.008 \\
D           &1.530 &-1.005  & 0.008 \\
G           & 1.314 & 0.067 & 0.06 \\
\hline \hline
\end{tabular}
\caption{F140W image positions in units of arcseconds with North up, East left coordinates (rotated relative to Figure 1).}
\end{table}

\begin{table*}
\centering
\scriptsize\begin{tabular}{llllll}
\hline
Filter & Date & $f_A/f_C$ & $f_B/f_C$ & $f_D/f_C$ & Reference \\
\hline
$u'$       & 9/2007 & 2.36$\pm 0.07$ & $0.93 \pm 0.04 $& $1.03 \pm 0.05$ &\citet{Blackburne++11} \\
CIV   & 9/2002 & 1.40$^{\rm{a}}$ & 1.05 &  0.770 & \citet{Wisotzki++03} \\
$g'$       & 9/2007 & 1.95$\pm 0.1 $ & $0.96 \pm 0.08 $& $0.9 \pm 0.07$ &\citet{Blackburne++11} \\
$V$        & 8-9/2009&  $1.37\pm 0.07$ & $0.97\pm 0.04 $& $0.79 \pm 0.04$ &\citet{Ricci++11} \\
F555W & 8/2003 &  $1.84\pm 0.1$ & $1.08 \pm 0.09 $& $0.95 \pm 0.06$ &\citet{Kochanek++06} \\
$R$       &  10/2007 & 1.89$\pm 0.05 $ & $0.99 \pm 0.04 $& $0.88 \pm 0.05$ &\citet{Blackburne++11} \\
$R$     &  10-12/2009 & $1.42\pm 0.08$ & $0.93 \pm 0.04 $& $0.75 \pm 0.04$ &\citet{Courbin++11} \\
$i'$       & 9/2007 &  $1.67\pm 0.09$ & $0.95\pm 0.07$& $0.77 \pm 0.06$ &\citet{Blackburne++11} \\
F814W & 8/2003 &  $1.69\pm 0.04$ & $1.12 \pm 0.04 $& $0.82\pm 0.03$ &\citet{Kochanek++06} \\
$z'$       & 9/2007 &  $1.64\pm 0.05$ & $0.96\pm 0.04$& $0.8 \pm 0.04$ &\citet{Blackburne++11} \\
$J$       & 9/2007 &  $1.57 \pm 0.09$ &$ 0.99 \pm 0.08$ & $0.69 \pm 0.06$ & \citet{Blackburne++11}\\
H$\beta$    & 8/2015   &  1.36 $\pm0.04$ &	 1.00 $\pm0.04$   &0.78$\pm0.03$ &	this work \\
{[OIII]}    & 8/2015    &  0.97$\pm0.07$  & 0.98$\pm0.07$  &  0.66$\pm0.05$ &	this work \\
F160W & 8/2003 & $1.57 \pm 0.05$ &$ 1.0 \pm 0.03$ & $0.79 \pm 0.03$ & \citet{Kochanek++06} \\ 
F160W & 10/2012 & $1.30 \pm 0.05$ &$ 0.92 \pm 0.05$ & $0.66 \pm 0.05$ & \citet{Wong++16} \\ 
$K_s$    & 9/2007 & $1.39 \pm 0.03$ &$ 0.98 \pm 0.03$ & $0.77 \pm 0.02$ & \citet{Blackburne++11} \\ 
$K$        & 8/2008 & $1.84 \pm 0.1$ &$ 1.37 \pm 0.08$ & $0.74 \pm 0.06$ & \citet{Fadely++11} \\ 
5 GHz    & 11/2012 & $1 \pm 0.1$ &$ 0.8 \pm 0.07$ & $0.47 \pm 0.07$ & \citet{Jackson++15} \\ 

\hline \hline
\end{tabular}
\caption{Subset of flux ratio measurements of HE0435-1223 from the literature, selected to represent variation across wavelength and with time, plotted in Figure 3. 
${^{\rm{a}}}$ Formal measurement uncertainties for \citet{Wisotzki++03} less than $0.1$\% and dominated by unknown systematics.}
\end{table*}

\label{lastpage}
\bsp
\end{document}